\documentclass[a4paper,11pt]{article}
\pdfoutput=1 

\usepackage{jinstpub} 

\usepackage[utf8]{inputenc}
\usepackage{graphicx}
\usepackage{subcaption}
\usepackage{array}
\usepackage{stackengine}

\title{Quenching factor measurements of sodium nuclear recoils in NaI:Tl determined by spectrum fitting}

\author[a,1]{L.~J.~Bignell\note{Corresponding author.},}
\author[b]{I.~Mahmood,} 
\author[b]{F.~Nuti,}
\author[a]{G.~J.~Lane,}
\author[a]{A.~Akber,}
\author[b]{E.~Barberio,}
\author[b]{T.~Baroncelli,}
\author[a]{B.~Coombes,}
\author[b]{W.~Dix,}
\author[a]{J.~T.~H.~Dowie,}
\author[a]{T.~Eriksen,}
\author[a]{M.~S.~M.~Gerathy,}
\author[a]{T.~J.~Gray,}
\author[a]{B.~P.~McCormick,}
\author[a]{A.~J.~Mitchell,}
\author[c]{M.~S.~Rahman,}
\author[b]{F.~Scutti,}
\author[a]{N.~J.~Spinks,}
\author[a]{A.~E.~Stuchbery,}
\author[c]{H.~Timmers,}
\author[b]{P.~Urquijo,}
\author[a]{L.~Wang,}
\author[a]{Y.~Y.~Zhong,}
\author[b]{and M.~Zurowski}

\affiliation[a]{Department of Nuclear Physics, The Australian National University, Canberra, ACT 2601, Australia}
\affiliation[b]{School of Physics, University of Melbourne, Melbourne, Victoria 3010, Australia}
\affiliation[c]{School of Science, University of New South Wales, Canberra, ACT 2610, Australia}

\emailAdd{lindsey.bignell@anu.edu.au}
\abstract{We have performed measurements of sodium nuclear recoils in NaI:Tl crystals, following scattering by neutrons produced in a $^{7}$Li(p,n)$^{7}$Be reaction. Understanding the light output from such recoils, which is reduced relative to electrons of equivalent energy by the quenching factor, is critical to interpret dark matter experiments that search for nuclear scattering interactions. We have developed a spectrum-fitting methodology to extract the quenching factor from our measurements, and report quenching factors for nuclear recoil energies between 36 and 401 keV. Our results agree with other recent quenching factor measurements that use quasi-monoenergetic neutron sources. The new method will be applied in the future to the NaI:Tl crystals used in the SABRE experiment.}

\keywords{Scintillators, scintillation and light emission processes (solid, gas and liquid scintillators), Dark Matter detectors (WIMPs, axions, etc.), Detector modelling and simulations I (interaction of radiation with matter, interaction of photons with matter, interaction of hadrons with matter, etc), Instrumentation for heavy-ion accelerators}

\begin{document}

\maketitle

\section{Introduction}
Weakly Interacting Massive Particles (WIMPs) are a well-motivated dark matter candidate that are expected to interact with ordinary matter via nuclear scattering. There is a large and ongoing search for such interactions from galactic WIMPs in terrestrial detectors, with most experiments returning null results \cite{PDG2020}. A notable exception is the DAMA experiment, which for many years has observed a modulation in the count rate of low energy events in their NaI:Tl detector consistent with that expected from WIMP dark matter \cite{Bernabei2018}. This observation has motivated numerous other NaI:Tl-based searches that aim to independently measure the DAMA modulation \cite{Antonello2019, Adhikari2019, Amar2019}, though none have unambiguously confirmed or refuted this result to date.

Understanding the scintillation response of NaI:Tl to nuclear recoils is essential to interpret any potential dark matter signal. The light yield from nuclear recoils is reduced relative to electron recoils of the same energy due to energy losses to lattice excitations and additional non-radiative carrier de-excitation pathways in regions of high excitation density \cite{Hitachi2005}. Furthermore, in practice, detectors are calibrated using gamma-ray sources that produce electron recoils, so to set the correct energy scale for nuclear recoil interactions the quenching factor
\begin{equation}
    Q = \frac{L_{\rm NR}}{L_{\rm ER}}
\end{equation}
must be known. Here $L_{\rm NR}$ and $L_{\rm ER}$ are the light yields due to a nuclear recoil and electron recoil of equal energy, respectively.

The DAMA experiment has measured values of 0.3 and 0.09 for Na and I recoils, respectively, using a $^{252}$Cf source and a simulation of the expected distribution of recoils \cite{Bernabei1996}. More recent experiments using quasi-monoenergetic neutrons from deuteron-deuteron neutron generators and $^{7}$Li(p,n)$^{7}$Be reactions have indicated an energy dependence where the Na quenching factor decreases with energy, though there is some inconsistency between these results \cite{Collar2013, Xu2015, stiegler2017, Joo2019}. This study provides an additional measurement of the quenching factor of Na recoils in a NaI:Tl crystal, and exploits a full spectrum fitting methodology as a means to improve the modelling of the recoil spectrum.

\section{Experiment}
\begin{figure}
    \centering
    \includegraphics[width=0.8\textwidth]{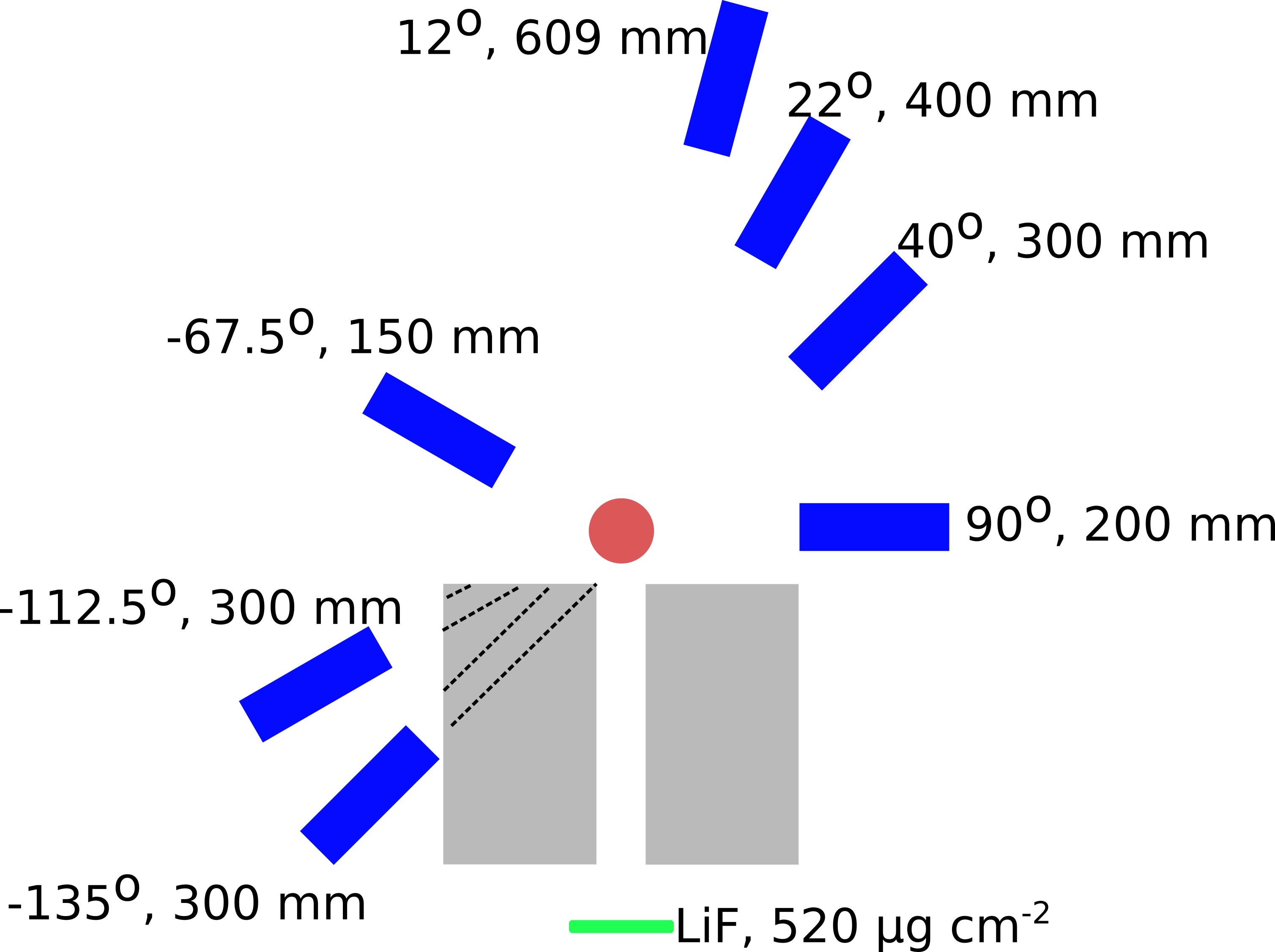}
    \caption{A schematic drawing of the nuclear recoil experiment (not to scale). Neutrons were produced in the tantalum-backed LiF target using $^{7}$Li(p,n)$^{7}$Be. These were collimated with a cylinder of HDPE (grey) onto the NaI:Tl (red) and the time-of-flight to events in the liquid scintillator detectors (blue), along with particle identification via pulse shape discrimination, were used to tag nuclear recoils. The polar angle with respect to the beam axis and the distance from the face of each liquid scintillator to the centre of the NaI:Tl is shown. The dashed lines indicate through-holes in the HDPE collimator.}
    \label{fig:Schematic}
\end{figure}

Figure \ref{fig:Schematic} illustrates the experimental setup. The choice of geometry was guided by a toy model that used cross-section data and kinematic calculations to predict coincidence rates, neutron times of flight, and nuclear recoil energies (including their spread) based on a simplified angular acceptance.

\subsection{NaI:Tl Detector}
An encapsulated 40 mm diameter, 40 mm height, cylindrical NaI:Tl crystal was employed to measure the nuclear recoils. The crystal was originally manufactured in the USSR circa 1988, and the growth method and thallium concentration are unknown. The crystal enclosure includes a transparent window, allowing readout with photomultipliers. Two Hamamatsu H11934-200 1 x 1 inch ultra-bialkali (43\% peak quantum efficiency) photomultipliers were coupled to the window using optical grease before wrapping with layers of teflon tape and black adhesive tape. The light yield from this setup was measured, using calibration sources and the peak charge from single photoelectrons, to be 2 PE/keV, which was lower than expected. We attribute this low sensitivity to a combination of poor optical coupling, owing to the mismatch between the crystal window and the PMT size, and a low intrinsic light yield from the crystal.

\subsection{LiF Target and Proton Beam}
A $520$~$\mu$g/cm$^{2}$ LiF foil on a 20~mg/cm$^{2}$ Ta backing was placed at the target location. This was irradiated by a pulsed beam of protons provided by the 14UD accelerator at the Australian National University's Heavy Ion Accelerator Facility \cite{Bignell2020}. A range of beam energies was chosen, sufficient to generate quasi-monoenergetic neutrons via $^{7}$Li(p,n)$^{7}$Be \cite{NeutronCS_LISKIEN197557, SIMLIT_FRIEDMAN2013117}. The target backing was sufficient to stop the primary proton beam, but had a negligible effect upon the neutrons. The pulsed beam width had a FWHM less than 2 nanoseconds, repeated every 747 ns. The beam current varied throughout the measurements, and also with beam energy, in a range between approximately 0.2 and 2 nA. Neutron production was confirmed at the time of measurement using the time difference spectrum between a BaF$_{2}$ scintillation detector placed close to the target and the beam pulse timing signal. This time spectrum exhibited a prompt peak from gammas emanating from the target and a delayed peak due to neutrons (section \ref{sec:syncdrift}). Two measurement campaigns were carried out, with proton beam energies of 2.44 MeV, 3 MeV, and 5.2 MeV in the first and 3 MeV and 6 MeV in the second.

Neutrons exited the vacuum chamber through a thinned aluminium vacuum flange. A cylindrical HDPE collimator was placed after the target so that the 25 mm aperture was collinear with the beam axis. The NaI:Tl detector was aligned with the aperture, 30 cm from the LiF target. This collimator included 25 mm diameter through-holes to allow the transport of neutrons to the backward-angle scattered neutron detectors.

\subsection{Liquid Scintillator Detectors}
Seven liquid scintillator detectors were placed at the angles and distances from the centre of the NaI:Tl crystal outlined in figure \ref{fig:Schematic} to tag scattered neutrons. 
The detectors were aligned using a theodolite, so the error in the mean angle is expected to be negligibly small.

The scintillation detectors were made with a diameter of 38 mm and a length of 15 cm. The length was chosen to give a balance between a sufficiently high macroscopic scattering cross-section (> 90\% for the neutrons traversing the cylindrical axis) and a small neutron time of flight over the length of the sensitive volume. The neutron time of flight over the liquid scintillator volume varied from 14 ns to 5 ns at the lowest and highest beam energies, respectively. The inner surfaces of the detectors were fabricated from 5 mm thick teflon, aside from a borosilicate glass window sealed with a Viton o-ring. The scintillation light was read out via a 1.5 inch diameter Hamamatsu H10828 photomultiplier with a super bialkali photocathode (35\% peak quantum efficiency), coupled to the detector window using optical grease. A thin aluminium housing covered the outer surfaces of the liquid scintillator detectors as light shielding, supplemented with duct tape and aluminium foil.

We used EJ-309, a low-hazard liquid scintillator with excellent particle identification capabilities, as the detection material. The scintillator was bubbled with nitrogen before filling the detectors in a nitrogen glove box. A small gas volume was included with the liquid scintillator to allow for thermal expansion.

\subsection{Data Acquisition}

An XIA Pixie16 digitizer \cite{XIA2020} was used to capture digitized waveform signals from all 9 detectors (7 liquid scintillators, 2 NaI:Tl channels) at 500 million samples per second with 12 bit resolution in both runs. A beam sync signal was generated as a logic signal in phase with the beam bunches. This was captured as a constant-fraction discriminator (CFD) time relative to the Pixie16 clock, not as a waveform. In the second run, the CFD timing signal from the BaF$_{2}$ neutron monitor was also captured and used to correct for the drift in the phase relationship between the beam timing signal and the arrival of the beam burst at the target (section \ref{sec:syncdrift}). The data acquisition system did not have a global trigger, instead each channel was digitized whenever the signal exceeded a threshold. The thresholds for the detector channels were set low enough to measure single photoelectrons.

To prevent excessive triggering, a signal over threshold in the liquid scintillator or NaI:Tl channels required a coincident signal in one of the NaI:Tl or liquid scintillator channels, respectively. An ideal trigger for the beam timing signal would require a coincident liquid scintillator and NaI:Tl channel. However, due to a limitation of our digitizer control implementation, we instead required a two-fold multiplicity of coincident signals amongst all of the liquid scintillator and NaI:Tl channels. This meant that events where both NaI:Tl channels triggered dominated the data stream, however these events could be readily rejected in later analysis. When it was acquired, the rate in the BaF$_{2}$ detector was low enough that self-triggering was sufficient.

\section{Signal Processing}
The digitized waveforms were processed to estimate the pulse arrival time, charge, and a particle identification metric for each detector channel.

\subsection{Trigger Time}
A significant amount of RF noise was observed in the NaI:Tl waveforms. To mitigate this noise and reduce the analysis threshold, the waveforms were digitally filtered using a Wiener filter. A Wiener filter requires an estimate of the power spectral density of the signal ($S$), and the noise ($N$), which together give an estimate of the signal to noise ratio in Fourier space. The filter is then defined as
\begin{equation*}
    H = \frac{S/N}{1+S/N}
\end{equation*}
so that frequencies where the signal-to-noise ratio is high are passed, and frequencies where it is low are attenuated.

Since noise was most problematic for low amplitude pulses, the filter was optimised for single photoelectron events. A population of single photoelectron waveforms was selected using a time over threshold cut. These were then aligned in time by up-sampling the waveforms and aligning to the maximum. The averaged power spectral density in the trigger region about the single photoelectron signal was used as an estimate of $S$, and the averaged power spectral density of the pre-trigger region was used as an estimate of $N$.

For the adopted analysis threshold of $\sim$1 photoelectron, signal processing without filtering led to inaccurate timing for a large number of events due to misidentifying noise in the pre-trigger region as the pulse arrival time. 
This can be quantified by the fraction of waveforms for which the pulse was determined to arrive more than 100 ns before the expected trigger time set by the digitizer's trigger delay. 
This fraction was 48\% without filtering and 0.3\% with the filter. A qualitative sampling of the remaining events that fell outside this time window revealed that most remaining events were due to pile-up, triggering on the tails of cosmic events, and similar undesirable events. We required the pulse arrival time as determined by our processing to arrive within a timing window relative to the start of the waveform to exclude these remaining events.

The liquid scintillator detectors did not exhibit the same level of electronic noise, and their timing was determined using a threshold on the unfiltered waveforms.

\subsection{Pulse Charge}
\begin{figure}[tbh]
    \begin{subfigure}[b]{0.5\textwidth}
        \centering
        \includegraphics[width=\textwidth]{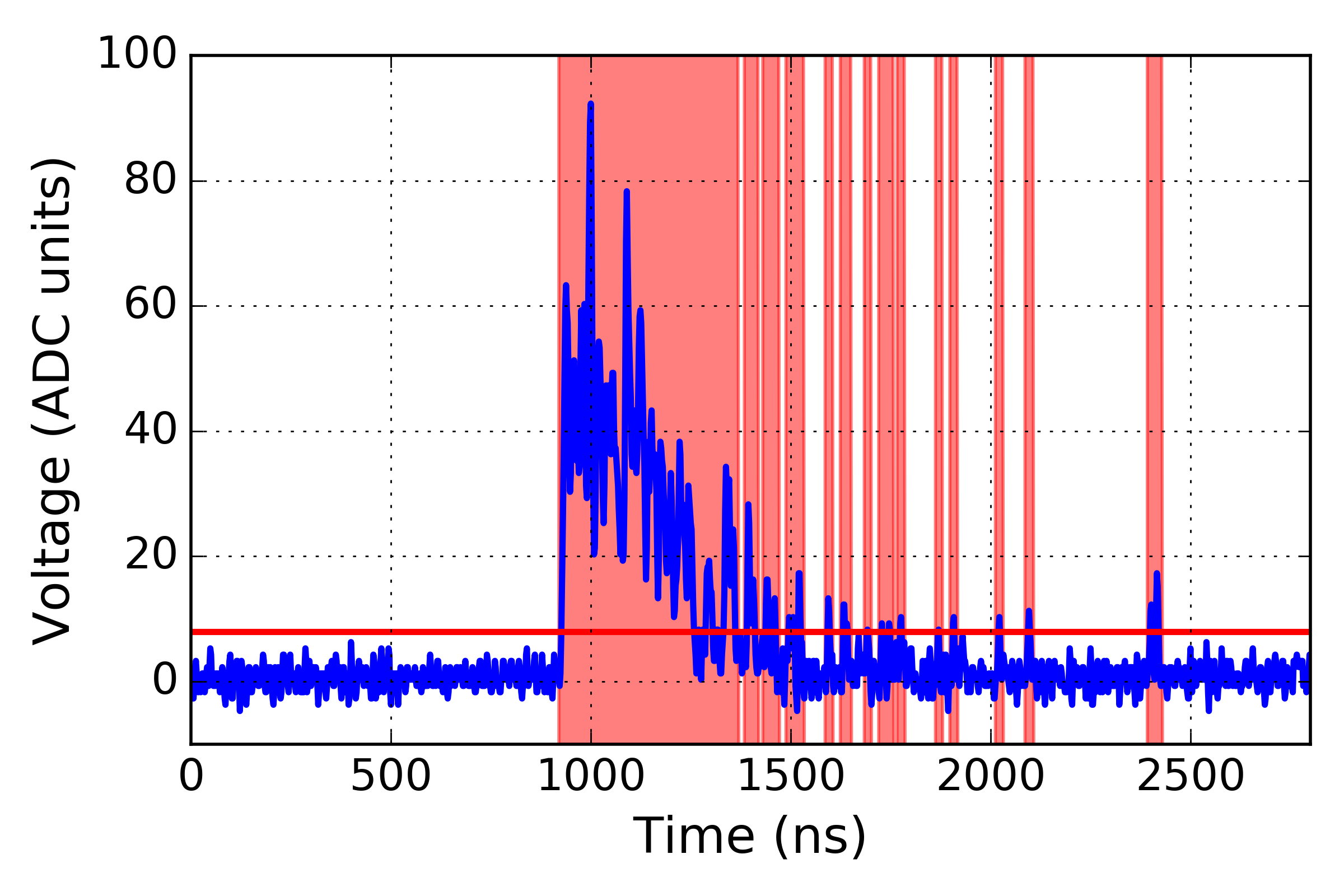}
        \caption{ }
        \label{fig:ChargeEstimateSchematic}
    \end{subfigure}
    \begin{subfigure}[b]{0.5\textwidth}
        \includegraphics[width=\textwidth]{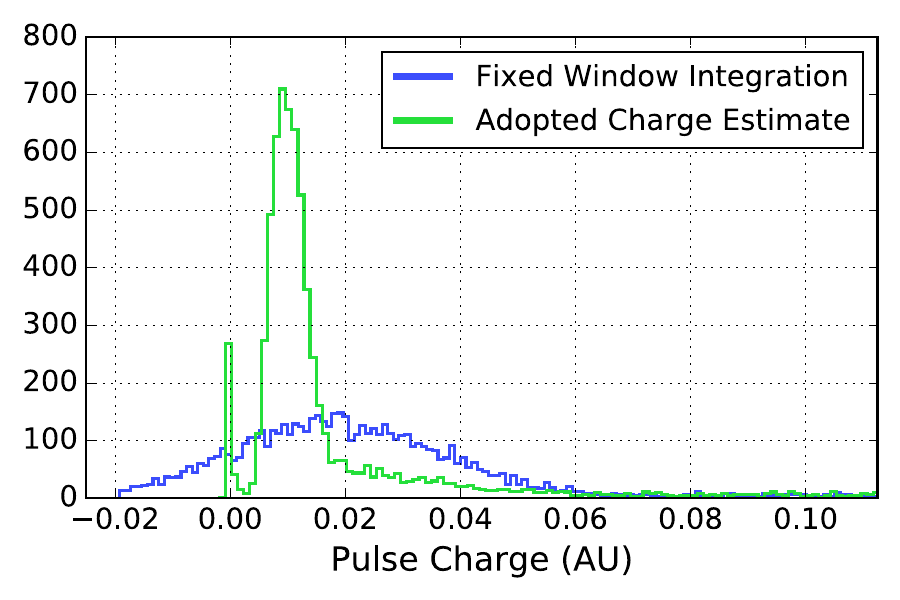}
        \caption{}
        \label{fig:ChargeComparison}
    \end{subfigure}
    \caption{(a) The pulse charge estimate integrated the charge only in regions where the waveform exceeded a $\sim$1 photoelectron threshold (red horizontal line), and extended the integral by 2 samples on either side of the crossing to allow for the rise and fall time of the photomultiplier response. The regions that are integrated for the waveform shown are highlighted in red. (b) The low charge performance of the adopted charge estimation is compared with that of a fixed integration window. The fixed integration window shows significantly worse single photoelectron performance.}
    \label{fig:ChargeEstimate}
\end{figure}

The low charge NaI:Tl events tend to be composed of single photoelectrons that are well-separated in time owing to the 250 ns NaI:Tl decay time, which is slow relative to the few ns signal width from a single photoelectron. Thus, a simple fixed integration approach performed poorly for low-charge NaI:Tl events and was unable to resolve the single photoelectron peak due to integrating the noise between well-separated single photoelectron events.

To avoid this, only regions around where the waveform exceeded a threshold were integrated. The threshold was chosen to be approximately 0.5 photoelectrons, to optimise the energy resolution. This algorithm is outlined schematically in figure \ref{fig:ChargeEstimateSchematic}. This approach performed much better than a fixed integration window at low charge, allowing single photoelectron events to be identified (figure \ref{fig:ChargeComparison}), while maintaining identical performance at high charge. This algorithm was adopted for all NaI:Tl energy estimates in this analysis. 

\subsection{Efficiency Correction}
\label{sec:eff}
\begin{figure}
    \centering
    \includegraphics[width=\textwidth]{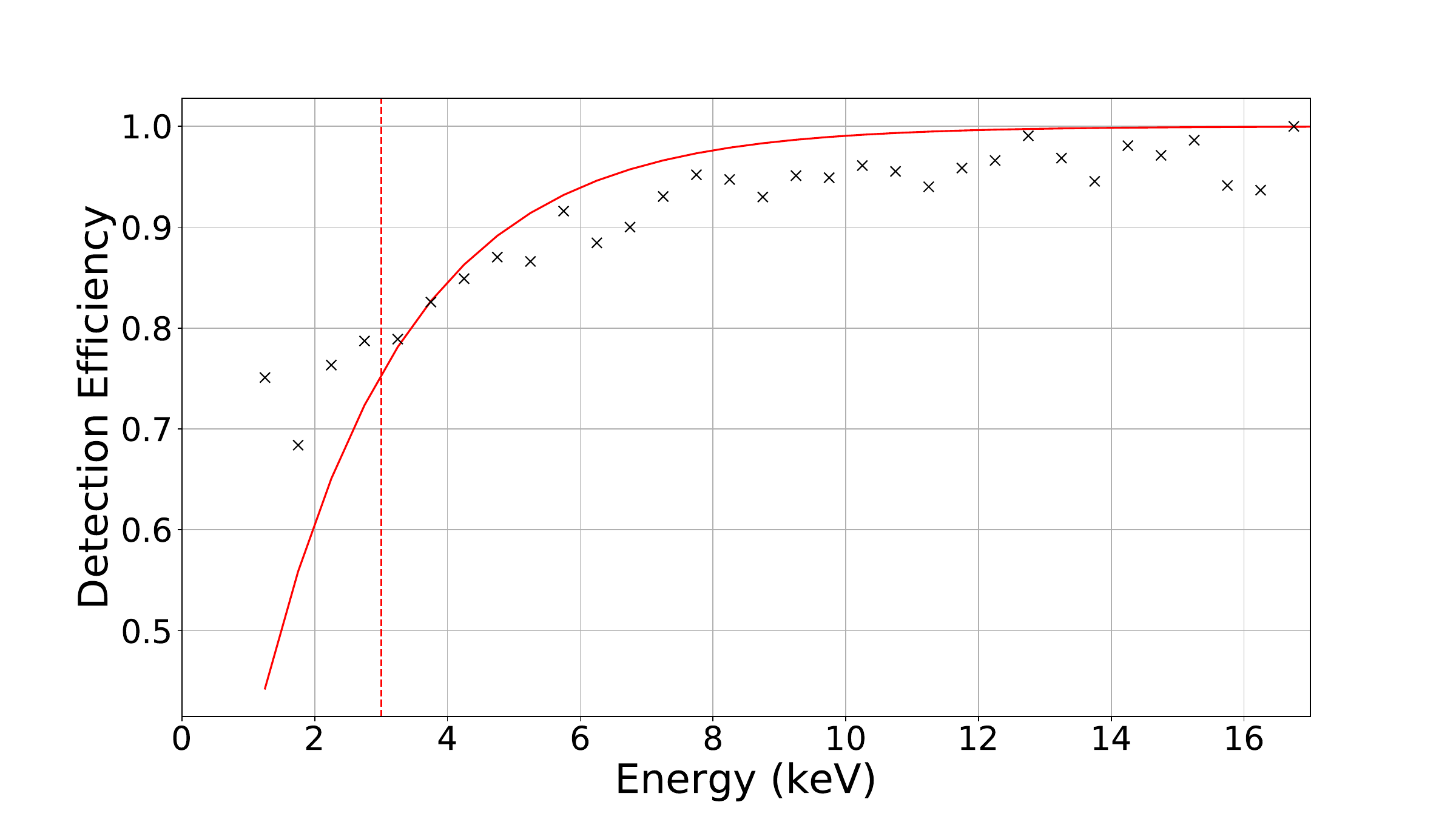}
    \caption{The measured energy-dependent detection efficiency, as well as its fit (see section \ref{sec:eff}). The fit values were used above the vertical dashed line.}
    \label{fig:deteff}
\end{figure}

The NaI:Tl detection efficiency drops below 1 at low energies due to there being few scintillation photons per event. Thus, the numbers of events measured near the threshold is an underestimation of the true spectrum, which can lead to incorrect predictions of the quenching factor \cite{Collar2010}. 
We have estimated the detection efficiency using coincidences between the two NaI:Tl photomultipliers; the efficiency of an individual channel is given by
\begin{equation}
    \epsilon_i = \frac{N_C}{N_j},
\end{equation}
where $N_C$ is the number of observed coincidences between channel $i$ and $j$ and $N_j$ is the number of counts in channel $j$ evaluated in an identical energy range. The coincident efficiency is the product of the two efficiencies and the efficiency of the union of the two channels is given by:
\begin{equation}
    \epsilon_U = \epsilon_1 + \epsilon_2 - \epsilon_c,
\end{equation}
where $\epsilon_c$ is the coincident efficiency. The union of all events passing the cuts was used for the analysis of our measurements, and the resulting spectra were corrected by an energy-dependent fit to this efficiency given by 
\begin{equation}
    f(E) = 1 - e^{-\lambda E}.
\end{equation}
This model function is motivated by the fact that the detection efficiency ought to be 1 minus the Poisson probability of detecting zero photoelectrons. The fit is given in figure \ref{fig:deteff}. The model fit is not perfect, and there are differences in the higher energy range that may stem from the difficulty of measuring unit efficiency with limited statistics. We imposed an analysis threshold at 3 keV in part to mitigate the degraded fit at low energies. The effect of the fit to the efficiency does not have a significant systematic effect on our quenching factor results: a reanalysis of the quenching factors without any efficiency correction resulted in a $\leq 0.5\%$ change to the quenching factor.

\subsection{Particle Identification}
\label{sec:particleID}
A well-known attribute of liquid scintillators is that the time distribution of the emitted scintillation light depends upon the stopping power of the particle that deposited energy in the material. This property can form the basis of particle identification.

\begin{figure}
    \centering
    \includegraphics[width=\textwidth]{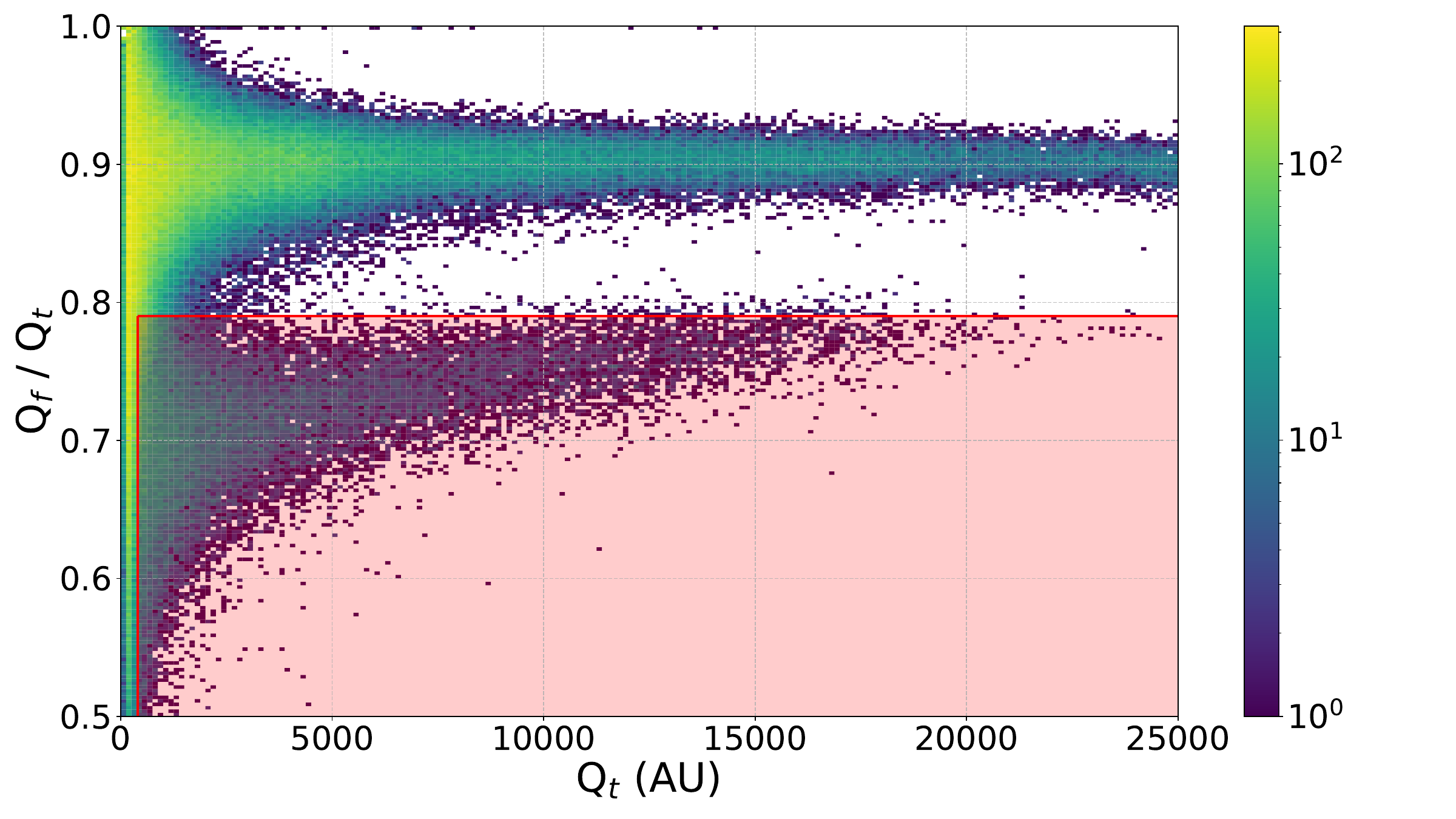}
    \caption{ Particle identification in the liquid scintillation detector at 67.5 degrees, with a proton beam energy of 5.2 MeV. See text for details of the definition of $Q_f$ and $Q_t$. The shaded red region indicates the selected nuclear recoil events. The upper band is from electron recoils, predominantly from $\gamma$ rays.}
    \label{fig:PSD}
\end{figure}

There are a number of techniques possible for particle identification. We used the charge comparison method \cite{ADAMS1978459}, with the metric Q$_f$/Q$_t$, where Q$_f$ is the charge in the first part of the pulse and Q$_t$ is the total pulse charge. Fixed integration windows about the pulse arrival time were used to estimate the charge. The free parameters for the algorithm are the integration lengths, which were determined using a parameter grid search to be optimal at 10 ns for $Q_f$ and 100 ns for $Q_t$. Figure \ref{fig:PSD} illustrates the excellent particle identification performance of the scintillator.

\subsection{Quality Cuts}
We applied a quality cut to all measurements to remove spurious noise events in the NaI:Tl photomultiplier channels. This charge asymmetry cut required that the difference in the proportion of charge detected between the two photomultipliers not be too great, such that
\begin{equation*}
    \frac{\left|Q_1 - Q_2\right|}{Q_1 + Q_2} < 0.5
\end{equation*}
where $Q_1$ and $Q_2$ are the charges measured in the two NaI:Tl photomultipliers.

We also applied a separate cut requiring that the two NaI:Tl signals arrive within 250 ns of each other.

\subsection{Beam Synchronisation Drift}
\label{sec:syncdrift}
\begin{figure}
    \begin{subfigure}[b]{0.5\textwidth}
        \centering
        \includegraphics[width=\textwidth]{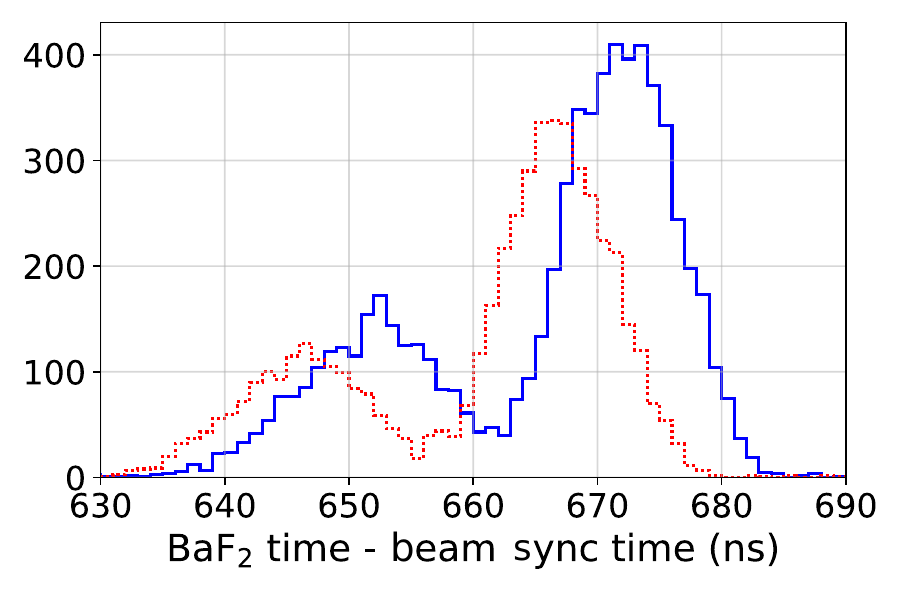}
        \caption{}
        \label{fig:SyncDriftTAC}
    \end{subfigure}
    \begin{subfigure}[b]{0.5\textwidth}
        \includegraphics[width=\textwidth]{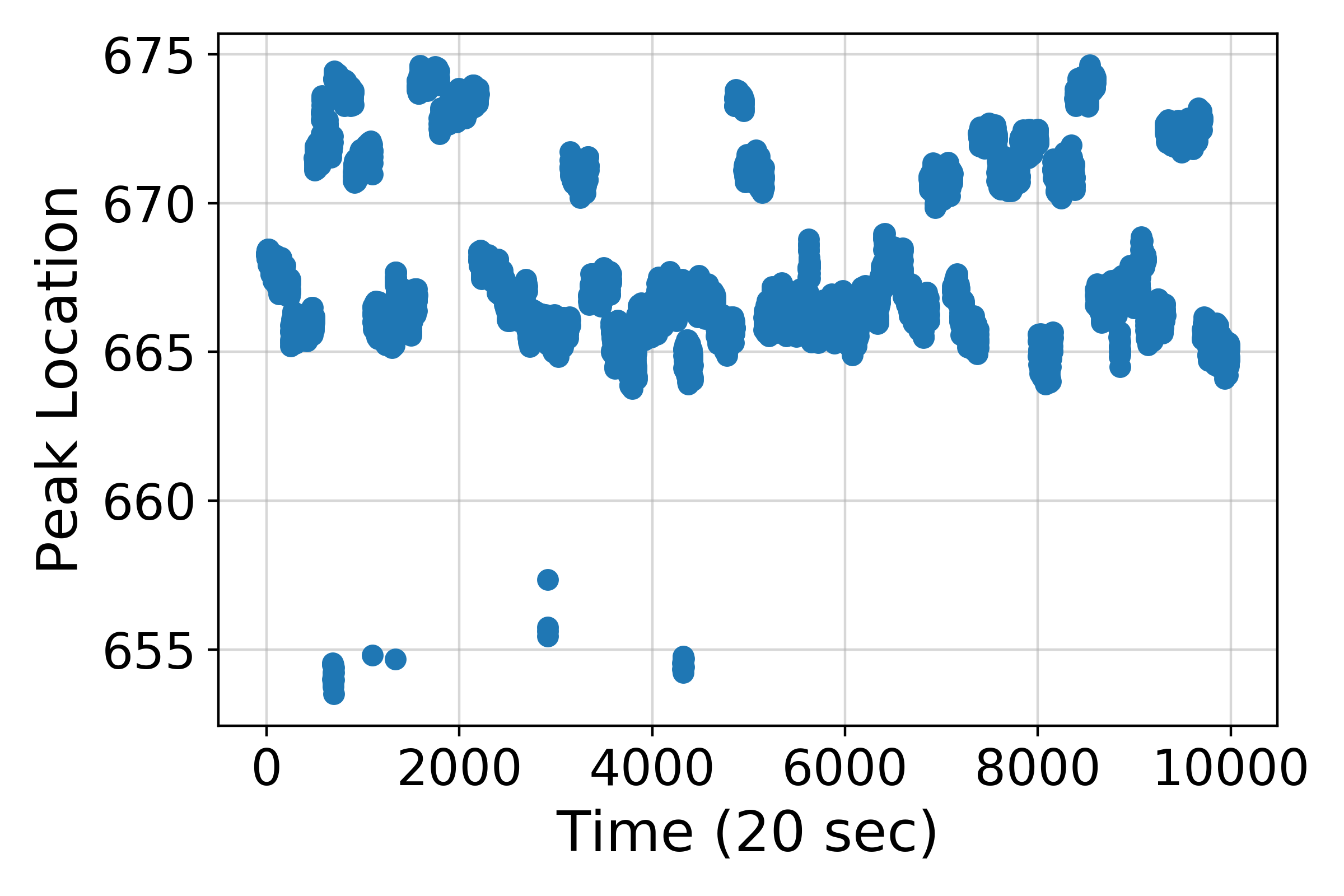}
        \caption{}
        \label{fig:SyncDriftFit}
    \end{subfigure}
    \caption{The drift of the beam pulse arrival time relative to the sync signal. (a) The time spectrum of prompt gamma rays and neutrons measured in the BaF$_2$ monitor, relative to the sync signal (so delayed events occur at earlier times). The larger and smaller peaks in each spectrum are due to prompt gamma rays and neutrons, respectively. The blue and red histograms represent arbitrary 20 second acquisitions within the 3 MeV beam measurement, and illustrate the level of drift encountered. (b) The fitted location of the beam arrival time across the entire 3 MeV measurement period.}
    \label{fig:SyncDrift}
\end{figure}

An analysis of the first measurement campaign revealed that the phase relationship between the beam sync signal and the arrival of the proton pulse at the target changed over time. In the second campaign this relationship was tracked by recording the timing signal from the BaF$_2$ signal, which yielded time spectra as shown in figure \ref{fig:SyncDriftTAC}. The two spectra, which represent 20 second slices of the run, show an obvious shift in the arrival time of the pulse relative to the beam sync signal.

To track this shift, we have performed fits to every analogous timing spectrum representing sequential 20 second slices of beam irradiation. The fit model used two Gaussian functions, and cuts were made to ensure the fit converged and achieved an acceptable goodness-of-fit. For time periods where the fits were not of sufficient quality, the mean of adjacent fits was used instead. In general, poor fits were only returned when the beam was not irradiating the target. The resulting distribution of the position of the prompt gamma signal during the 3 MeV beam measurement is shown in figure \ref{fig:SyncDriftFit}. There appears to be continuous small-scale variability as well as sudden step-like changes. The origins of these timing variations are not known, however they have been corrected for in our analysis.

\section{Results}
\subsection{Electron Recoil Analysis}

The combined energy estimate for all events was based on the summed charge contributions of both NaI:Tl photomultipliers in both nuclear and electron recoil measurements. The electron recoil energy scale was set using gamma ray sources: $^{133}$Ba, $^{137}$Cs, $^{152}$Eu, and $^{241}$Am. The spectral peaks were fit to a Gaussian function on a linearly interpolated background.

We corrected the measured charge in the NaI:Tl detector for the known non-linear electron recoil response \cite{Rooney1997, Khodyuk2010} to obtain a linearised electron-equivalent energy scale. A small non-linearity remained evident at high energies, which was accounted for by using a piecewise linear fit based on a linear calibration of the low-energy peaks below 80 keV and a separate calibration based on the high energy peaks above this energy.

A shift in the energy scale was evident during the irradiation, based on the movement of the location of the gamma ray associated with the first excited state of $^{127}$I at 57.6 keV. These could have been gain variations in the photomultipliers or due to differences in the spatial location of source calibration and beam events. The shifts were accounted for by tracking the position of this peak at the different beam energies and applying a uniform energy scale correction. Only the low-angle liquid scintillator detectors with iodine recoil energies less than 5 keV were used in the gain correction, so that the contribution from the iodine recoil energies did not strongly affect the peak location. (Note that the low quenching factor reduces the correction to approximately 250 eV.) The uncertainty of the corrections was limited by the number of inelastic scattering events collected at these angles, which translates to an additional 1\% uncertainty in the quenching factor results to account for the variability in the energy scale.

The energy resolution of the linearised NaI:Tl energies was fitted with a function proportional to the square root of the energy, and this function was used to smear the simulated spectra (section \ref{sec:sim}) with the experimental resolution.

\begin{figure}
    \begin{subfigure}[b]{0.5\textwidth}
        \centering
        \includegraphics[width=\textwidth]{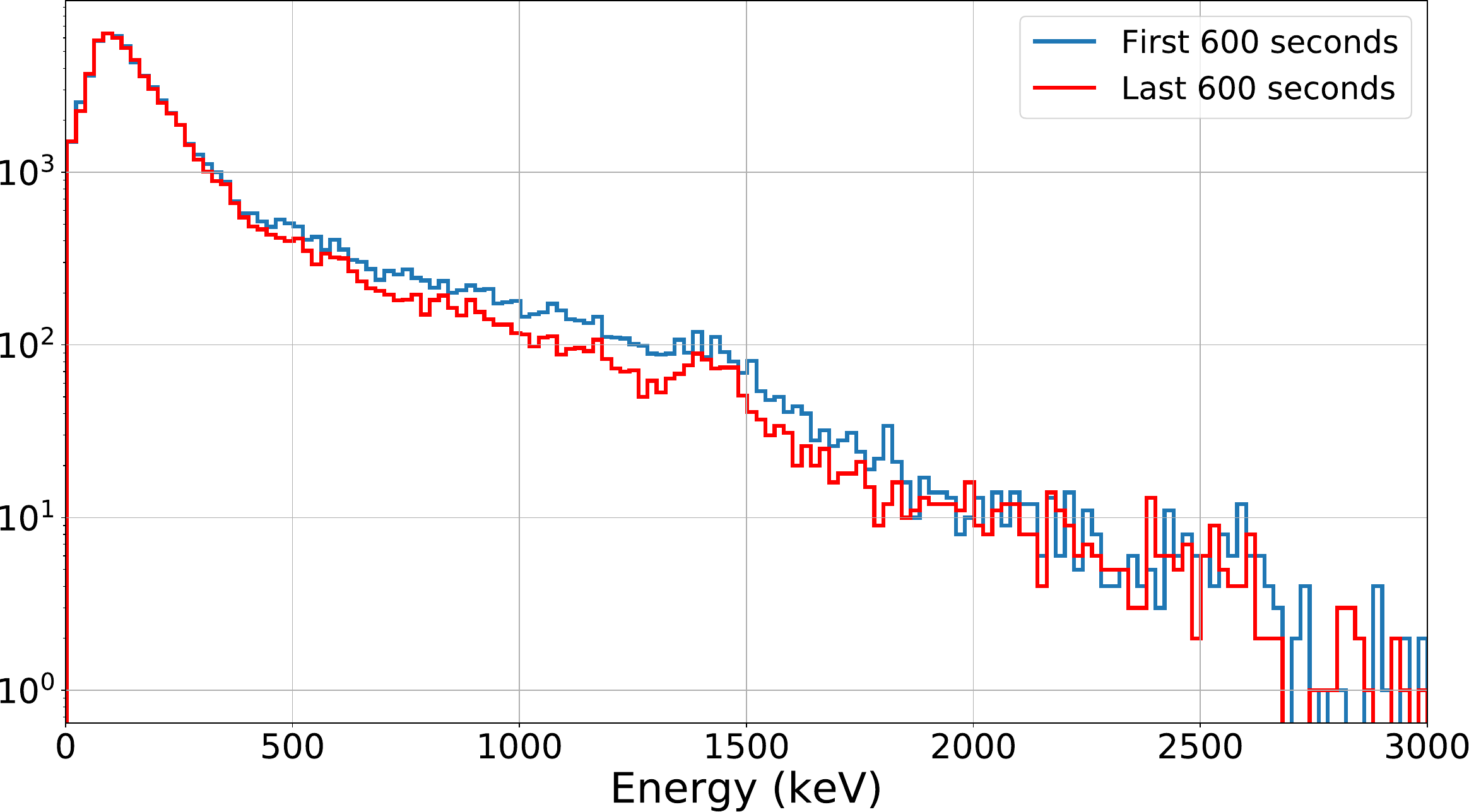}
        \caption{}
        \label{fig:BGspec}
    \end{subfigure}
    \begin{subfigure}[b]{0.5\textwidth}
        \includegraphics[width=\textwidth]{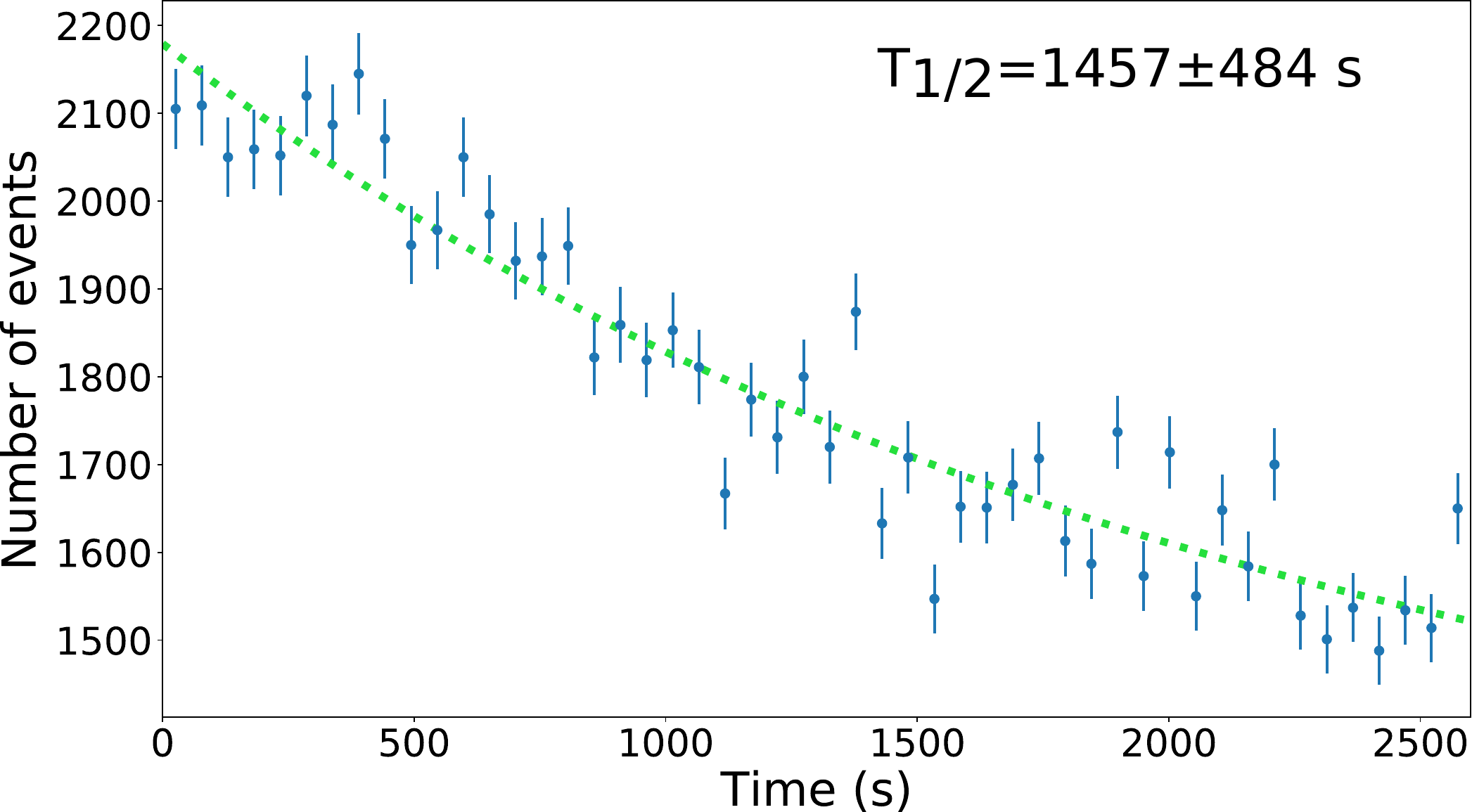}
        \caption{}
        \label{fig:BGdecay}
    \end{subfigure}
    \caption{Background measurements of the NaI:Tl crystal post-irradiation. (a) The energy spectrum in the first 600 and last 600 seconds of the 2600 second acquisition. The peak at 1461 keV is due to naturally occurring $^{40}$K background. (b) The time dependence of the event rate between 450 and 2000 keV, with a fitted exponential decay.}
    \label{fig:I128_decay}
\end{figure}

Following the irradiation, we captured background measurements of the NaI:Tl to investigate the activation of the crystal. The background rate varied throughout the 2600 second measurement, especially in the 450 keV to 2000 keV range (figure \ref{fig:BGspec}). A fit of the decay to an exponential function with constant background in this region gives a half-life of 1457 $\pm$ 484 seconds, which is consistent with the $^{128}$I half-life of 1499.4 $\pm$ 1.2 seconds. $^{128}$I is expected to be produced abundantly via neutron capture on $^{127}$I.

\subsection{Nuclear Recoil Analysis}
Nuclear recoils were selected using timing cuts and particle identification cuts in the liquid scintillator detectors.

\begin{figure}
    \centering
    \includegraphics[width=\textwidth]{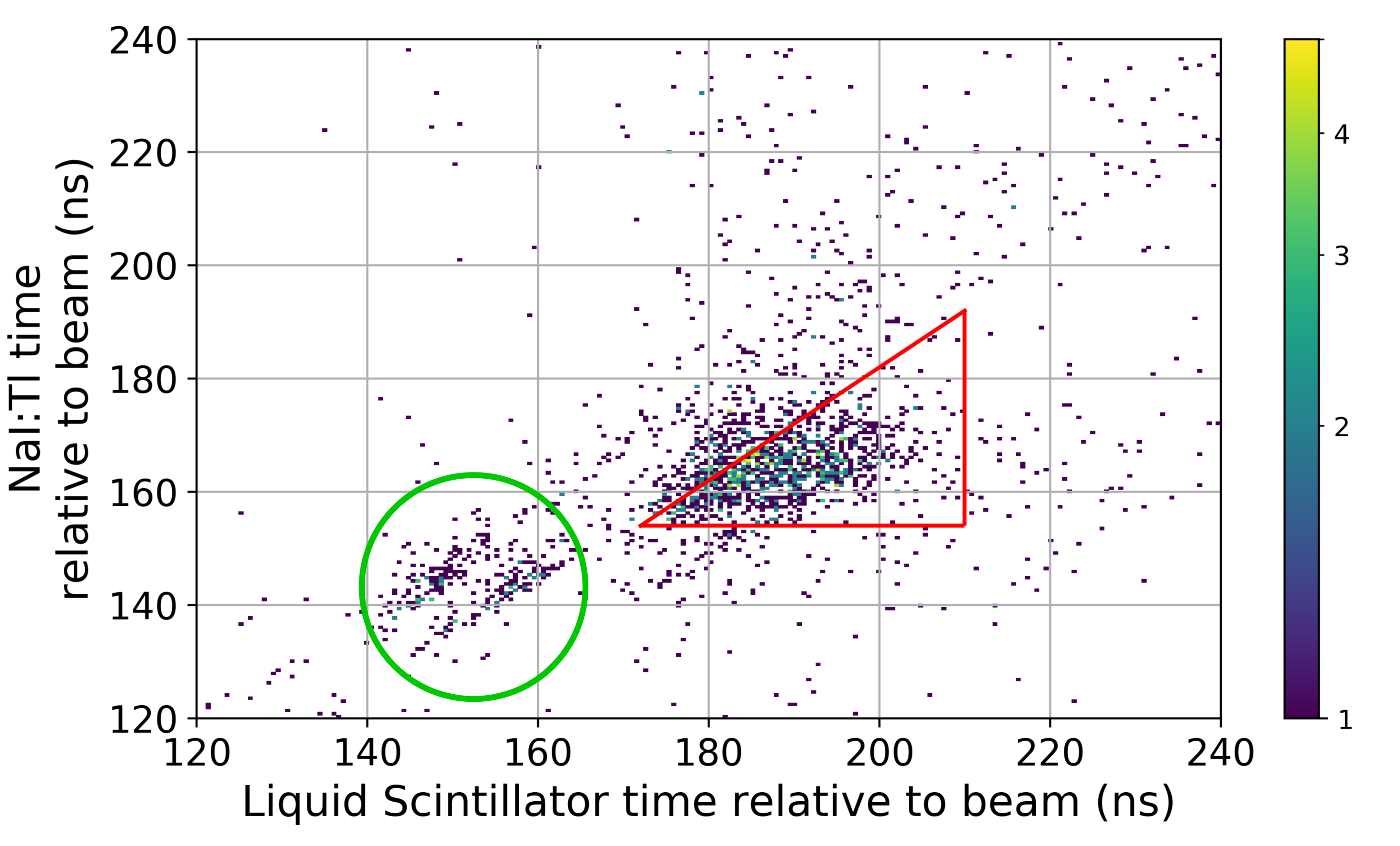}
    \caption{The time correlation of events in the NaI:Tl and nuclear recoil-like events in the 90$^\circ$ liquid scintillator detector, for the 3 MeV proton beam. Larger values are more delayed with respect to the beam arrival for both detectors. The green circle highlights the early beam-correlated events caused by gamma ray scattering, and the red triangular section illustrates the cuts used to select the later nuclear recoil events.}
    \label{fig:Timing}
\end{figure}

Figure \ref{fig:Timing} shows a time correlation plot (relative to the beam pulse time) for the 3 MeV proton beam dataset at 90$^\circ$ recoil angle. The population lying along the diagonal line represents fast coincidences from events such as Compton-scattered gamma rays and cosmic ray showers. The excess along this line (green circle, figure \ref{fig:Timing}) is due to prompt gamma ray interactions in both detectors. The timing distribution exhibits a bimodal spread due to a 10 ns offset that appeared in the digitizer, which is especially apparent in the prompt signal. This extra timing spread has required the use of stringent timing cuts, which are expected to affect the statistical power of the measurement but not the recoil spectrum. The nuclear recoil events are delayed in both the NaI:Tl detector and the liquid scintillator detectors, and the cut region used to select these events is shown on the figure. Given the measurement geometry used to generate the plot, the neutron arrival time should be delayed by $\sim19$~ns and $\sim36$~ns relative to the prompt gamma burst for the NaI:Tl and liquid scintillator detectors, respectively. These predicted times are consistent with the observed delays in figure \ref{fig:Timing}). Analogous cuts were made for all detectors and beam energies.

\section{Simulation}
\label{sec:sim}

\subsection{Neutron Beam Generation}

We used the Geant4 toolkit, version 10.5 \cite{GEANT4_AGOSTINELLI2003250}, to simulate the neutron interactions within the experimental geometry. In order to properly simulate the low energy neutrons generated for this experiment, we used the "QGSP\_BIC\_HP" physics list. This list uses nuclear data to transport neutrons with energies less than $\sim$20 MeV with high precision. Range cuts for neutrons and protons were reduced to zero in order to invoke the Geant4 processes needed to accurately simulate low energy behaviour. Primary neutrons were emitted in a 1 mm radius beam spot from the centre of the LiF target.

The energies and momenta of the neutrons were calculated using SimLiT \cite{SIMLIT_FRIEDMAN2013117}. SimLiT is a specialized Monte Carlo computational tool designed to calculate the neutron energy and angle distributions produced from $^{7}$Li(p,n)$^{7}$Be and $^{7}$Li(p,n)$^{7}$Be$^{*}$ reactions for given proton beam and target parameters. SimLiT natively supports proton energies above the $^{7}$Li(p,n)$^{7}$Be reaction threshold up to 3.5 MeV. This is lower than some of the energies used for our experiment. Therefore, we have extended the SimLiT energy range using neutron production cross sections from Liskien \textit{et al.} \cite{NeutronCS_LISKIEN197557} and LiF stopping powers from SRIM \cite{SRIM_ZIEGLER20101818}.

Due to neutron emission from the $^{7}$Be excited state, the proton beam produces two quasi-monoenergetic, forward-directed neutron beams. Neutrons from the $^{7}$Li(p,n)$^{7}$Be$^{*}$ reaction make up 0.8\%, 20\%, and 10\% of the total neutrons produced for the 2.44, 3 and 5.2 MeV proton beams respectively. Hence the $^{7}$Li(p,n)$^{7}$Be$^{*}$ reaction contributes a non-negligible portion of the total neutrons produced, for all but the lowest energy. 

\subsection{Recoil Energies}

\begin{figure}
    \centering
    \includegraphics[width=0.97\textwidth]{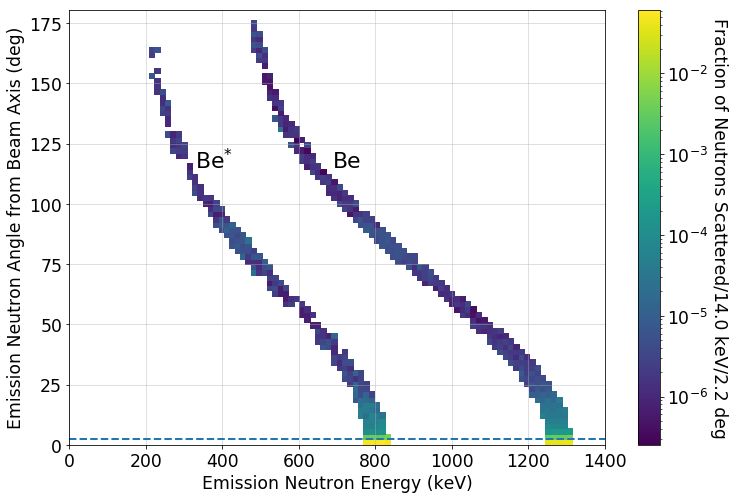}
    \caption{Fraction of total emitted neutrons that scatter in NaI detector and proceed to the scintillator placed at 67.5 degrees from the beamline. 73\% of neutrons are below the angle of acceptance at 2.2 degrees; shown in blue. Events in the left group (labelled Be$^{*}$) are neutrons produced from the $^{7}$Li(p,n)$^{7}$Be$^{*}$ reaction, while in the right group (labelled Be) neutrons are produced from the $^{7}$Li(p,n)$^{7}$Be reaction}
    \label{fig:SimEmissRatioSpec67p5_3MeV}
\end{figure}

We simulated the unquenched recoil spectra by tagging neutrons that deposit energy in the NaI:Tl and exit at a polar angle subtended by the front face of one of the liquid scintillator detectors used in our geometry. Figure \ref{fig:SimEmissRatioSpec67p5_3MeV} shows the fraction of the simulated neutrons that scattered in the NaI detector towards the scintillator placed at 67.5 degrees from the beam line for the initial 3 MeV proton energy; this represents 0.007\% of the simulated events. Most of these events, approximately 73\%, involve neutrons that travel straight into the NaI detector.
However, the remaining 27\% are neutrons that interact with other materials (predominantly the collimator) prior to reaching the NaI detector. Thus, they can have significantly different energies, as shown by the spread of events above the blue dashed line in figure \ref{fig:SimEmissRatioSpec67p5_3MeV}. Of these, neutrons that are not emitted in the forward direction towards the collimator can reach the detector by scattering off beamline and vacuum fittings surrounding the target. Of the neutrons that scatter into the 67.5 degree scintillator, 7\% are from the $^{7}$Li(p,n)$^{7}$Be$^{*}$ reaction. 

\begin{figure}
    \begin{subfigure}[b]{0.5\textwidth}
        \centering
        \includegraphics[width=\textwidth]{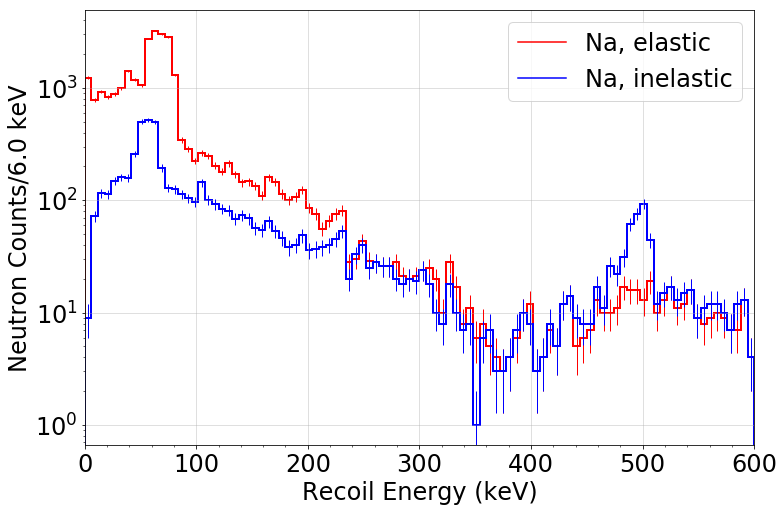}
        \caption{}
        \label{fig:SimRecoilSpec67p5_3MeV_Na}
    \end{subfigure}
    \begin{subfigure}[b]{0.5\textwidth}
        \includegraphics[width=\textwidth]{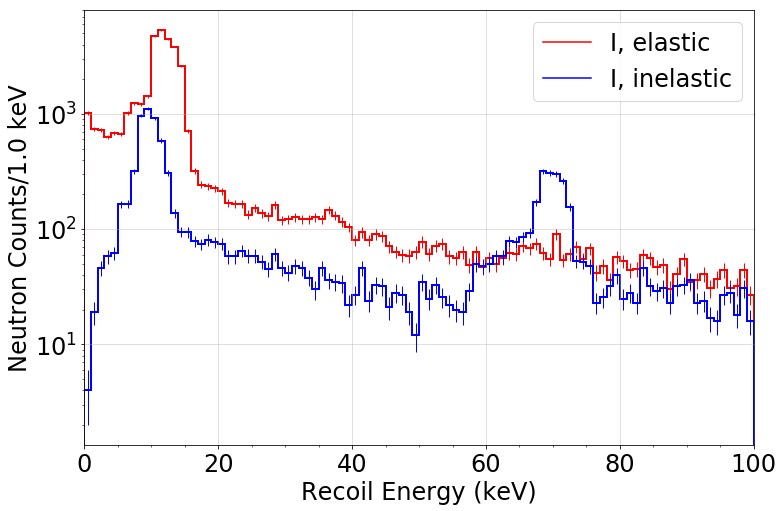}
        \caption{}
        \label{fig:SimRecoilSpec67p5_3MeV_I}
    \end{subfigure}
    \caption{Simulation of the unquenched energy deposited in the NaI detector due to neutrons that scatter to the angle subtended by the liquid scintillator placed at 67.5 degrees from the beamline, for the 3 MeV proton beam energy. (a) and (b) show the Na and I spectra, respectively. The red and blue lines show events where the neutron scatters elastically or inelastically, respectively. We have applied an energy threshold of 0.1 keV to remove thermal neutron events. The error bars show the standard Poisson error. Each of the spectra exhibits peaks that correspond to single scattering events. For the inelastic spectra, the peaks at lower energy correspond to events where the gamma ray escapes the detector.}
    \label{fig:SimRecoilSpec67p5_3MeV}
\end{figure}

Figure \ref{fig:SimRecoilSpec67p5_3MeV} shows the energy spectrum associated with the neutrons that were scattered into the liquid scintillator placed at 67.5 degrees from the beam line. The non-resonant backgrounds in figures \ref{fig:SimRecoilSpec67p5_3MeV_Na} and \ref{fig:SimRecoilSpec67p5_3MeV_I} are due to multiple neutron scattering, which makes up 23.8\% of events. Smaller peaks seen to the left of their larger counterparts (for example the small elastic Na peak centred at $\sim$40 keV before the larger one centred at $\sim$60 keV) in figure  \ref{fig:SimRecoilSpec67p5_3MeV_Na} are due to neutrons from the $^{7}$Li(p,n)$^{7}$Be$^{*}$ reaction. The elastic Na and I peaks in the low energy region <10 keV, are mainly due to events where neutrons lose energy in the collimator or aluminium casing before scattering in the detector. This was determined by comparing simulations where the collimator and aluminium casing were removed from the geometry.

\section{Quenching Factor Analysis}

Previous quasi-monoenergetic neutron studies of the quenching factor in NaI:Tl have fit the nuclear recoil peaks with Gaussian functions to extract the quenching factor. In our analysis, we have decided to fit the spectrum of simulated results to the experimentally measured recoil spectra. This approach allowed us to leverage the information contained in the inelastic recoil peaks, and also helped to incorporate complicating effects such as multiple scattering within the NaI:Tl, scattering from the collimator, and lower energy neutron emissions from the excited $^{7}$Be state. A comparative analysis involving fits with a simple Gaussian fit to the visible Na recoil peaks and linear background gave quenching factor results that differed from the spectrum fit analysis by between 7\% and 43\%. The fitted peak location also depended upon the choice of the fit range. Furthermore, not all recoil spectra could easily be fitted with a Gaussian. This is especially true for the higher beam energy data, where the elastic scattering peak was broadened by lower energy components associated with emissions from the excited $^{7}$Be state and inelastic scattering events where the gamma ray escaped the NaI:Tl.

\subsection{Spectrum Fit}
We applied multiplicative parameters to the simulated Na and I recoil energies to model the quenching factors. These parameters were adjusted so as to minimise the negative log-likelihood of the resulting simulated distribution, given the measured data. 

\begin{figure}
    \centering
    \includegraphics[width=\textwidth]{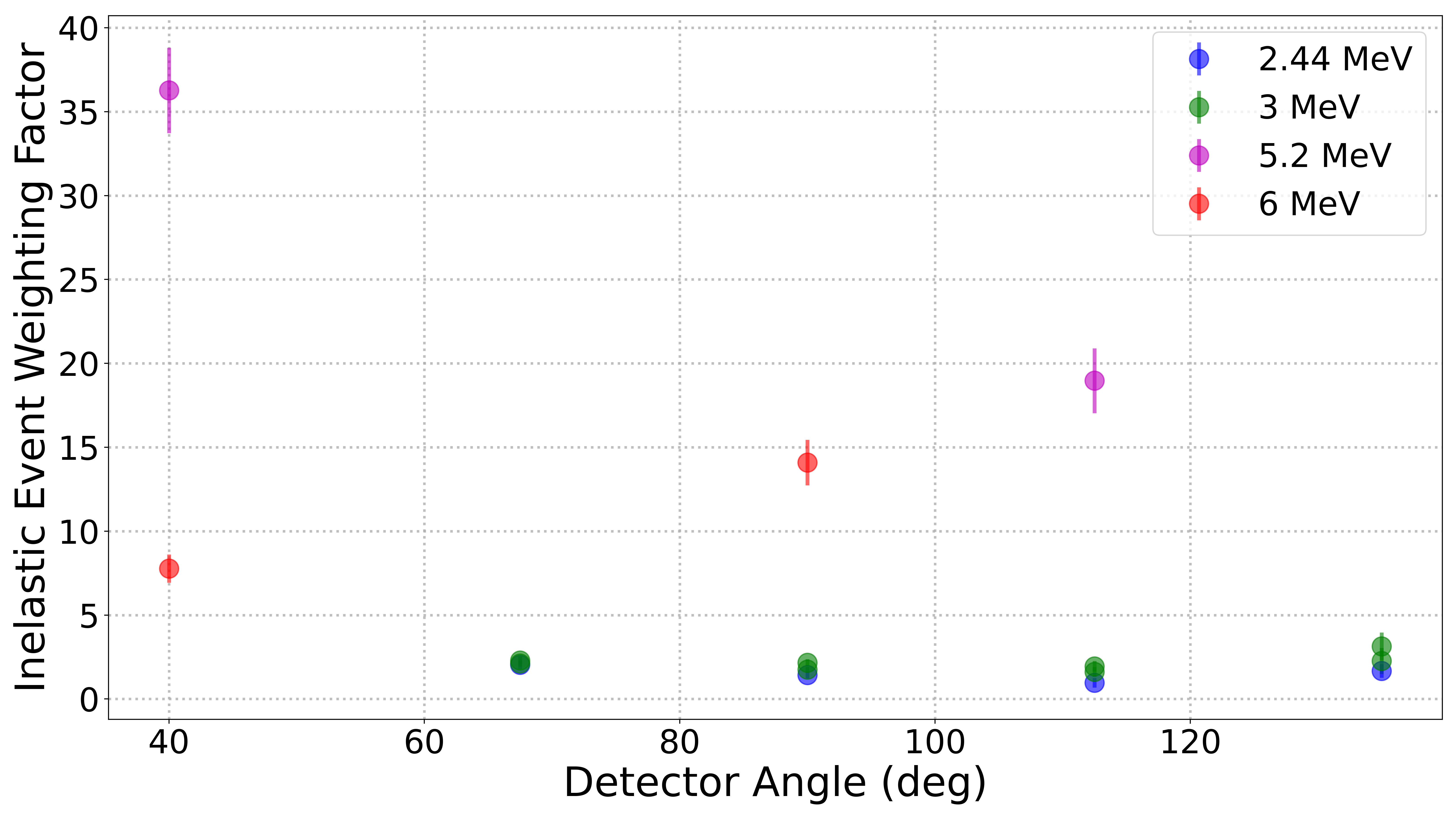}
    \caption{The event weighting factors for the inelastic $^{127}$I excitation. Only the fits used in the final analysis are shown. The different series denote the different proton beam energies.}
    \label{fig:InelWgts}
\end{figure}

An important feature in the recoil spectra is the peak at $\sim$58 keV from the inelastic excitation of $^{127}$I. The energy deposit from such events is the sum of the 57.6 keV gamma ray and the recoiling nuclear energy deposit. Since the I recoil energies were low due to the experimental kinematics, and the I quenching factor is low, the location of this peak did not move during our measurements. However, this feature was consistently under-predicted by the Geant4 model at all neutron energies, which led to poor fits. To mitigate this effect, a third parameter was introduced in the likelihood model which weighted these inelastic scattering events. Figure \ref{fig:InelWgts} shows the fitted inelastic scattering weights, which were broadly consistent with being near 2 for the 2.44 MeV and 3 MeV beam energies, and higher for the higher beam energies.

The fitted detector resolution was applied to the simulated spectra to allow comparison with the experimental results. Our early analysis of the resolution smearing showed that smearing the quenched nuclear recoil energies tended to under-predict the widths of the associated features in the spectrum. Instead we have applied the energy resolution to the unquenched nuclear recoil energies; this gives a better fit to our measurements.

\begin{figure}
    \centering
    \includegraphics[width=0.8\textwidth]{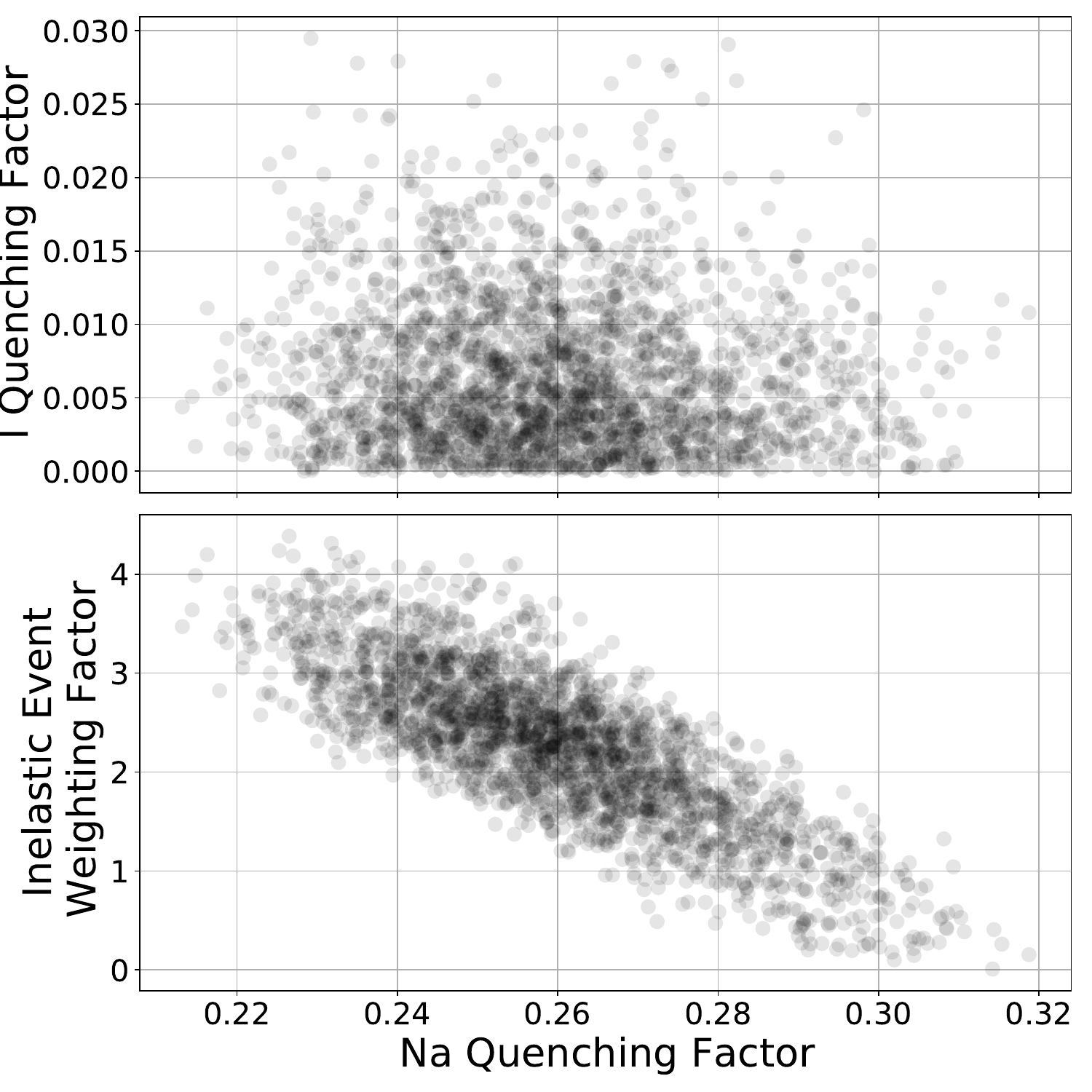}
    \caption{Every Markov Chain Monte Carlo sample for the fit to the NaI:Tl recoil spectrum observed in coincidence with the 135$^\circ$ liquid scintillator with a 3 MeV proton beam energy. 
    10000 Monte Carlo trials were used in the sampling. The distribution of samples obtained by this fit were used as an estimator of the posterior distribution of the fit parameters.}
    \label{fig:MCMCsampling}
\end{figure}

Once a maximum likelihood parameter estimate was made, we sampled via Markov Chain Monte Carlo in order to assess the posterior distribution of quenching factors and inelastic peak biasing allowed by the data. Figure \ref{fig:MCMCsampling} demonstrates this sampling for a single beam energy and liquid scintillator detector. Typical fits showed only weak correlations between the fit parameters, with the strongest correlation between the Na quenching factor and the peak biasing shown in the figure, with a correlation coefficient of 0.8. This correlation arose from a trade-off between the location of the Na elastic recoil peak and the inelastic I peak intensity, when the peaks were close in energy.

Recoil spectra were excluded from our analysis where the fitted quenching factors predicted recoil peaks below the 3 keV analysis threshold, since there is less informative data available to properly constrain the fit. Fits where the Na recoil peak coincided with the 57.6 keV inelastic excitation peak were also excluded, since this degeneracy also led to poor fitting constraints. The highest energy nuclear recoil distributions above the maximum fitted energy of 401 keV were also excluded, since these gave poor fits to the simulated spectra -- the Na elastic recoil peak tended to be broader than expected from simulation, and this flat spectrum did not adequately constrain the fit parameters.

\subsection{Results}

\begin{figure}
    \begin{subfigure}[b]{0.5\textwidth}
        \centering
        \phantomcaption
        \stackinset{l}{0.3in}{b}{1.4in}{(\thesubfigure)}{\includegraphics[width=\textwidth]{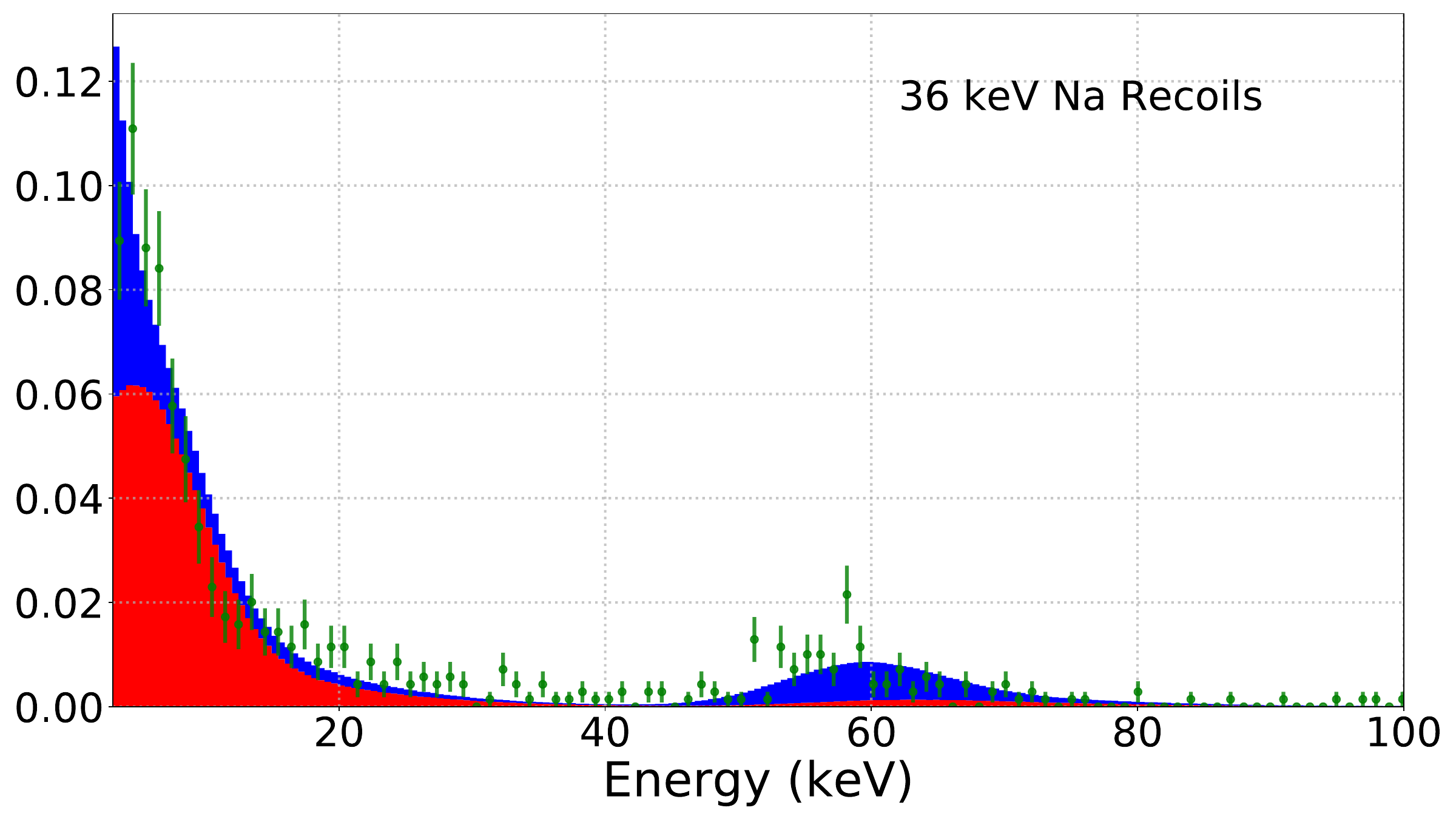}}
        \label{fig:fit2_4}
    \end{subfigure}
    \begin{subfigure}[b]{0.5\textwidth}
        \phantomcaption
        \stackinset{l}{0.3in}{b}{1.4in}{(\thesubfigure)}{\includegraphics[width=\textwidth]{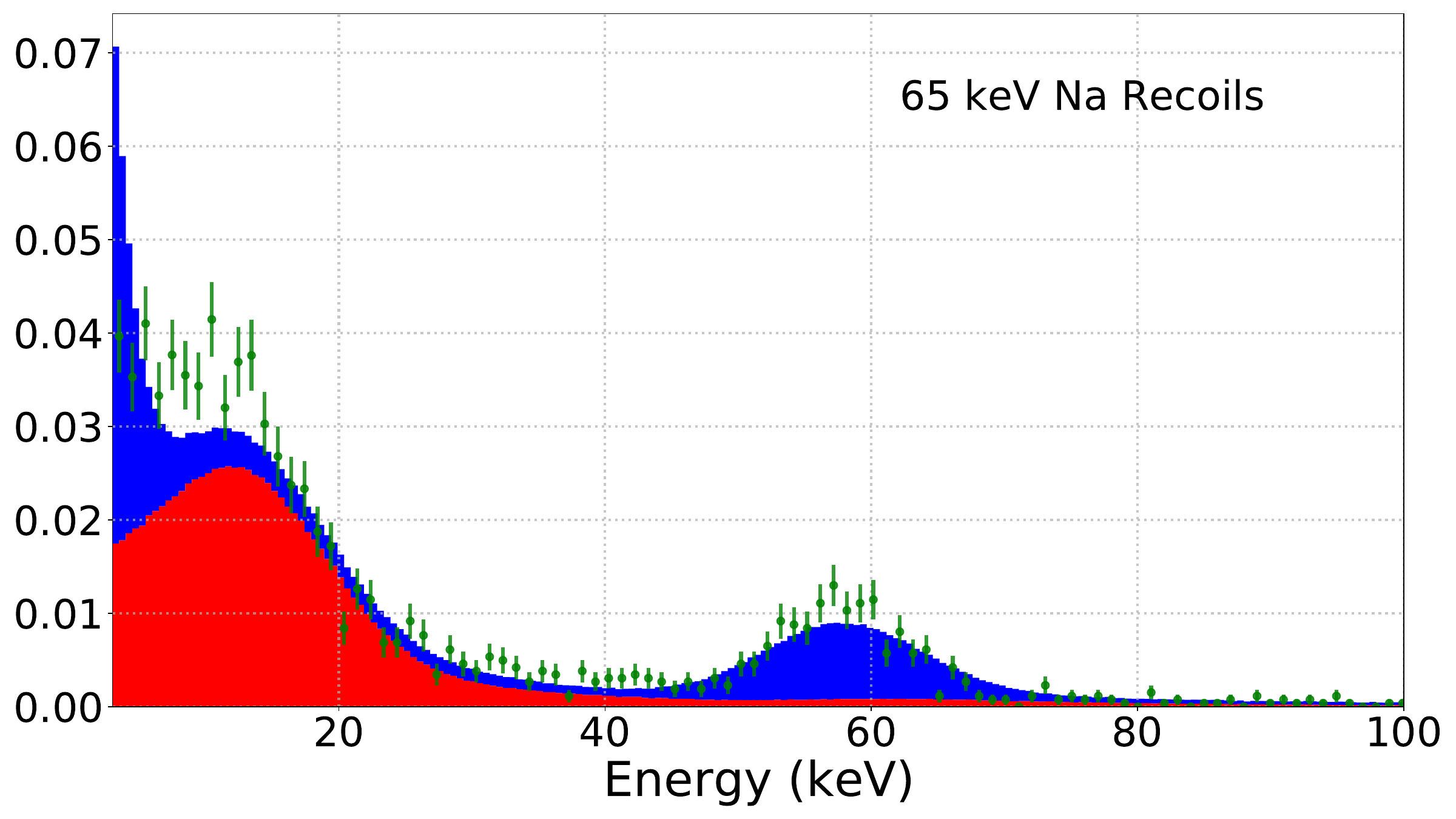}}
        \label{fig:fit3_4}
    \end{subfigure}
    \begin{subfigure}[b]{0.5\textwidth}
        \centering
        \phantomcaption
        \stackinset{l}{0.3in}{b}{1.4in}{(\thesubfigure)}{\includegraphics[width=\textwidth]{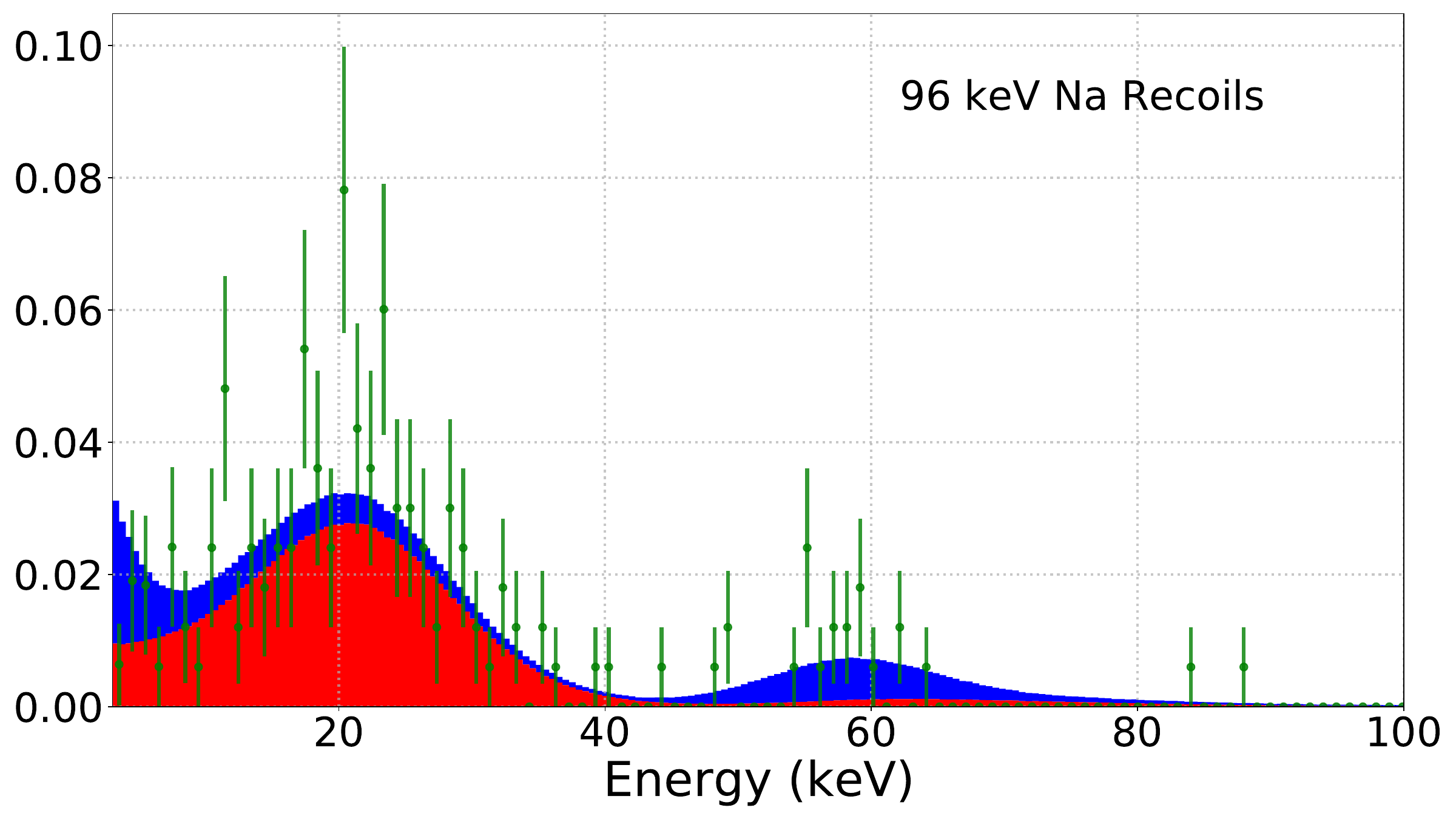}}
        \label{fig:fit2_6}
    \end{subfigure}
    \begin{subfigure}[b]{0.5\textwidth}
        \phantomcaption
        \stackinset{l}{0.3in}{b}{1.4in}{(\thesubfigure)}{\includegraphics[width=\textwidth]{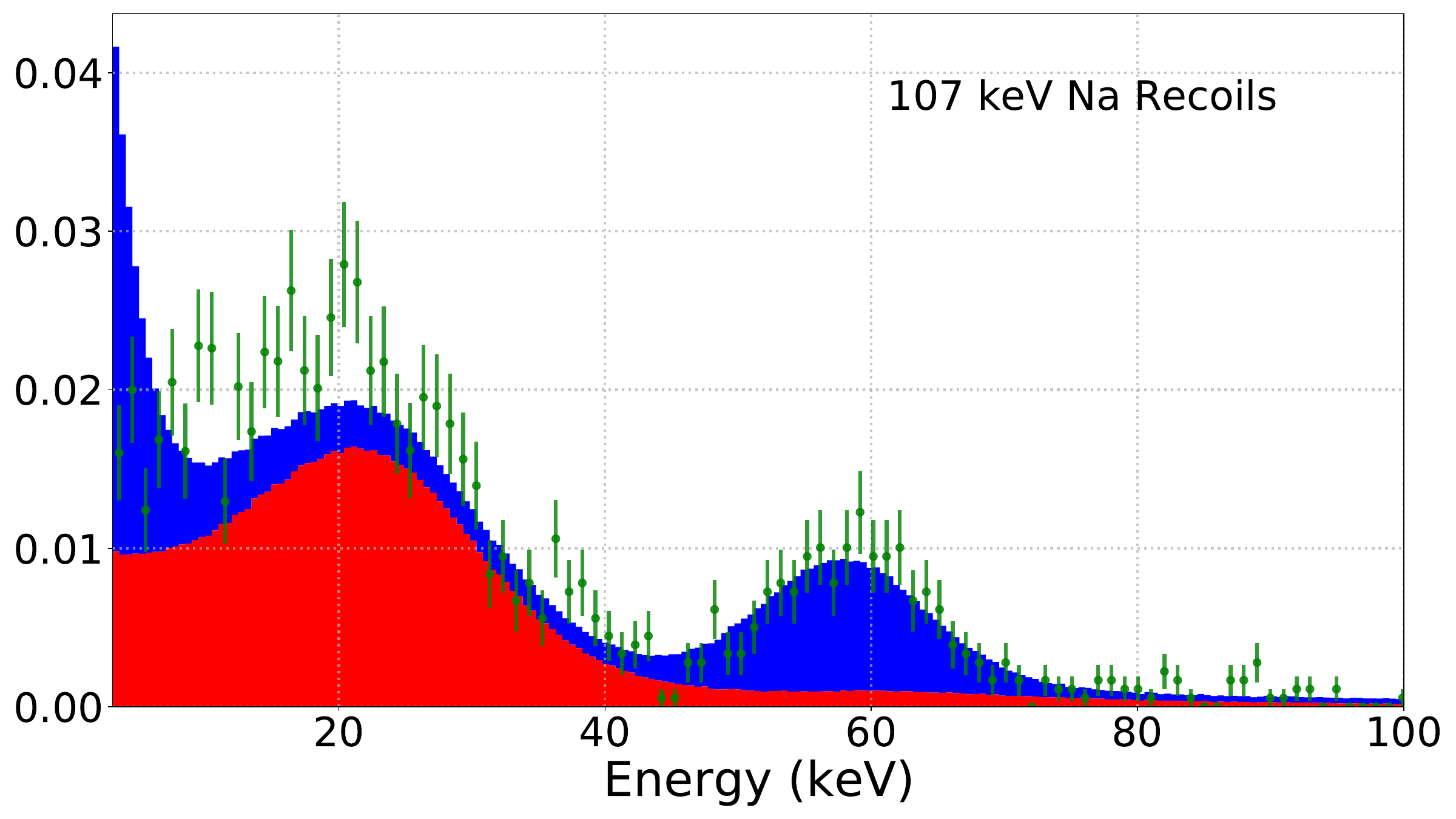}}
        \label{fig:fit3_3}
    \end{subfigure}
    \begin{subfigure}[b]{0.5\textwidth}
        \centering
        \phantomcaption
        \stackinset{l}{0.3in}{b}{1.4in}{(\thesubfigure)}{\includegraphics[width=\textwidth]{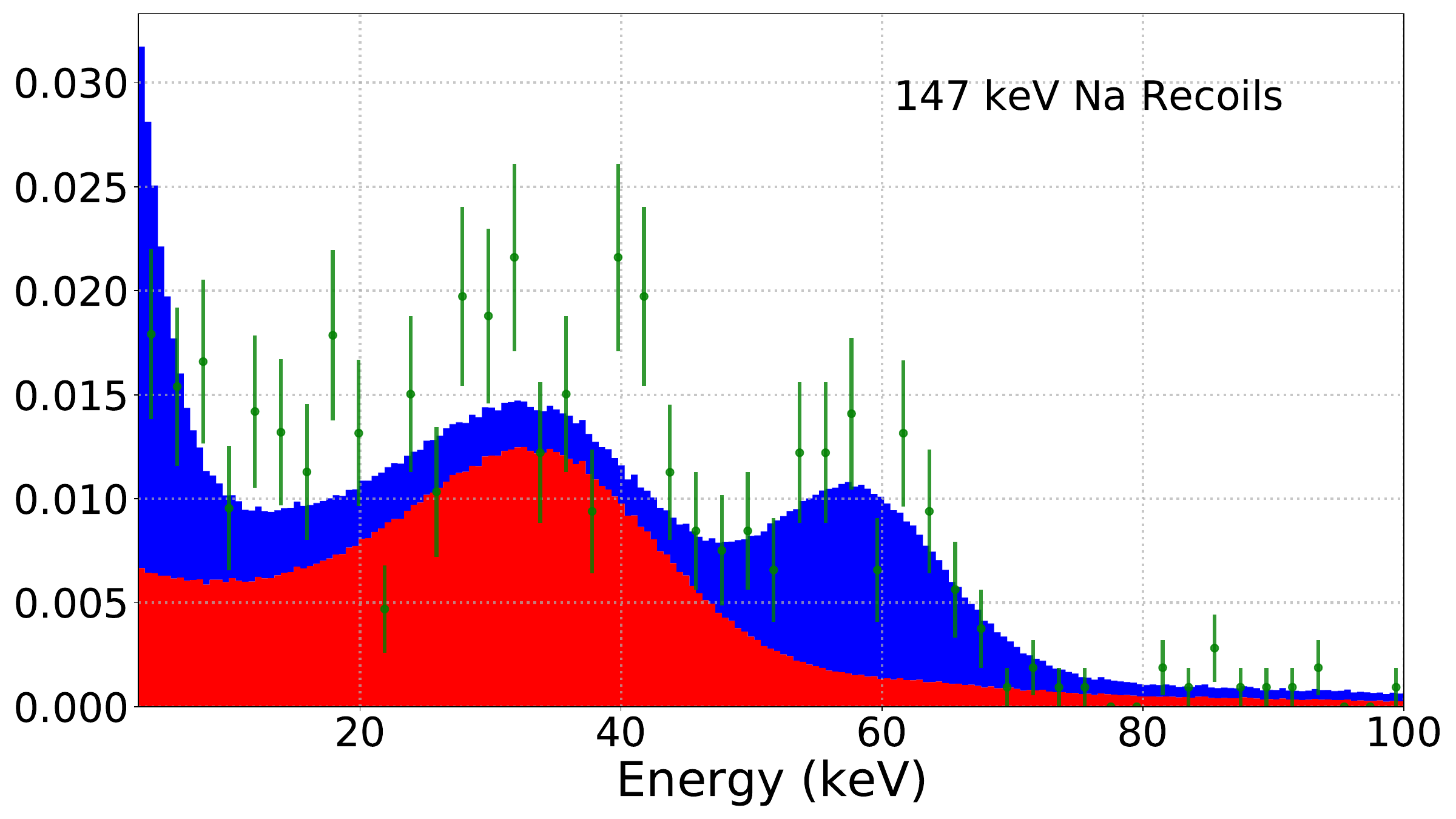}}
        \label{fig:fit3_5}
    \end{subfigure}
    \begin{subfigure}[b]{0.5\textwidth}
        \phantomcaption
        \stackinset{l}{0.3in}{b}{1.4in}{(\thesubfigure)}{\includegraphics[width=\textwidth]{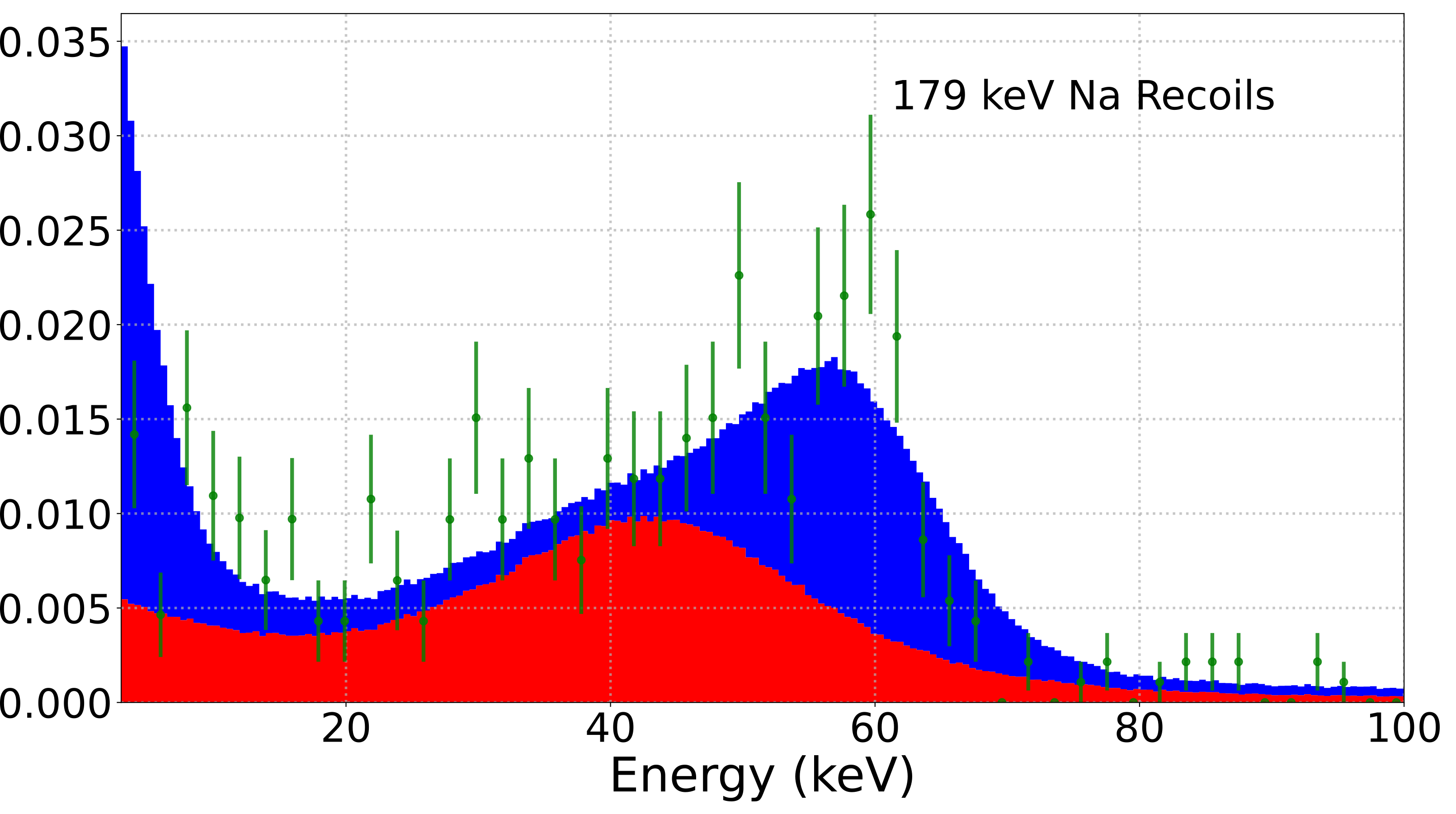}}
        \label{fig:fit3_6}
    \end{subfigure}
    \begin{subfigure}[b]{0.5\textwidth}
        \phantomcaption
        \stackinset{l}{0.3in}{b}{1.4in}{(\thesubfigure)}{\includegraphics[width=\textwidth]{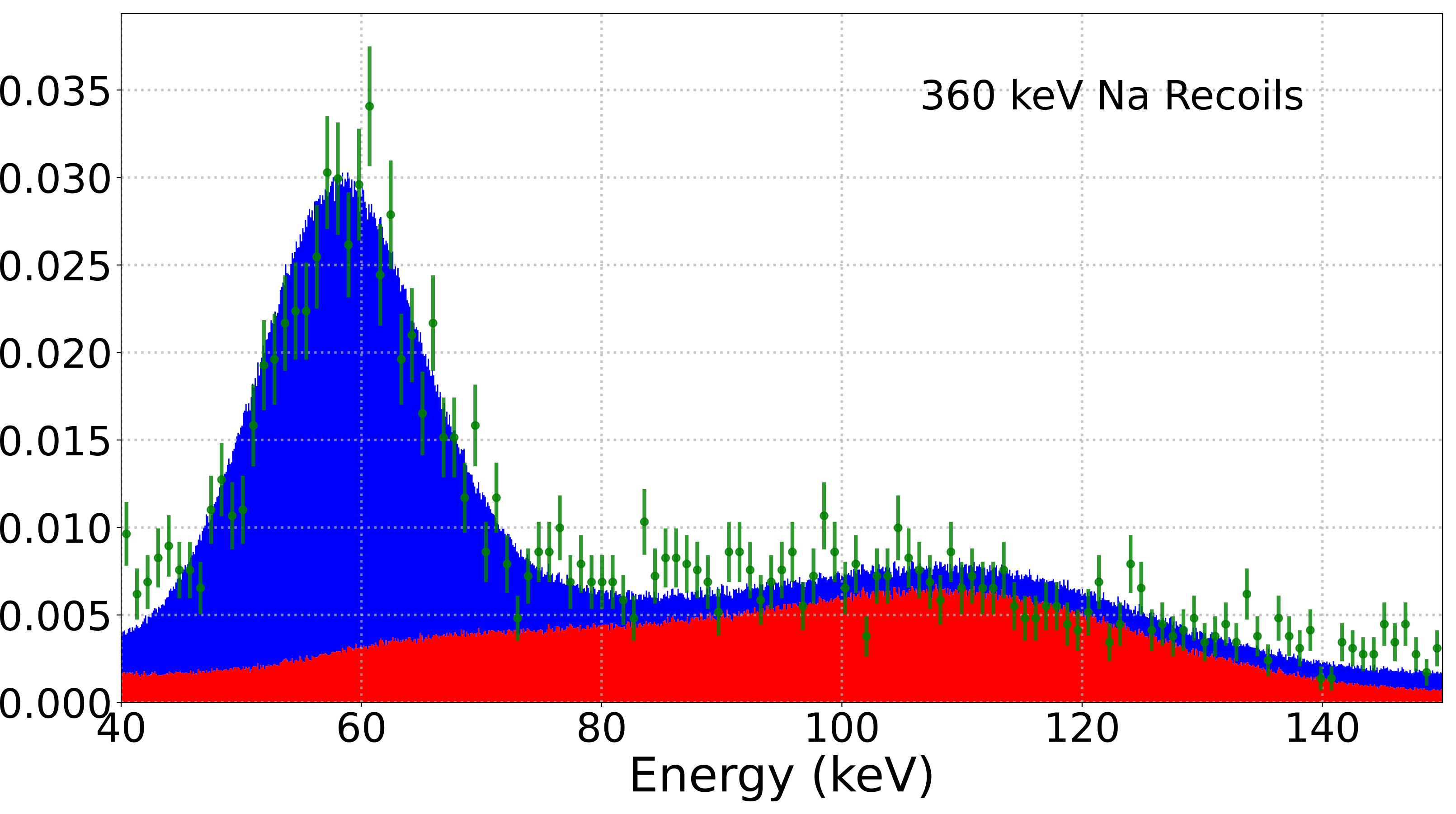}}
        \label{fig:fit6_3}
    \end{subfigure}
    \begin{subfigure}[b]{0.5\textwidth}
        \centering
        \phantomcaption
        \stackinset{l}{0.3in}{b}{1.4in}{(\thesubfigure)}{\includegraphics[width=\textwidth]{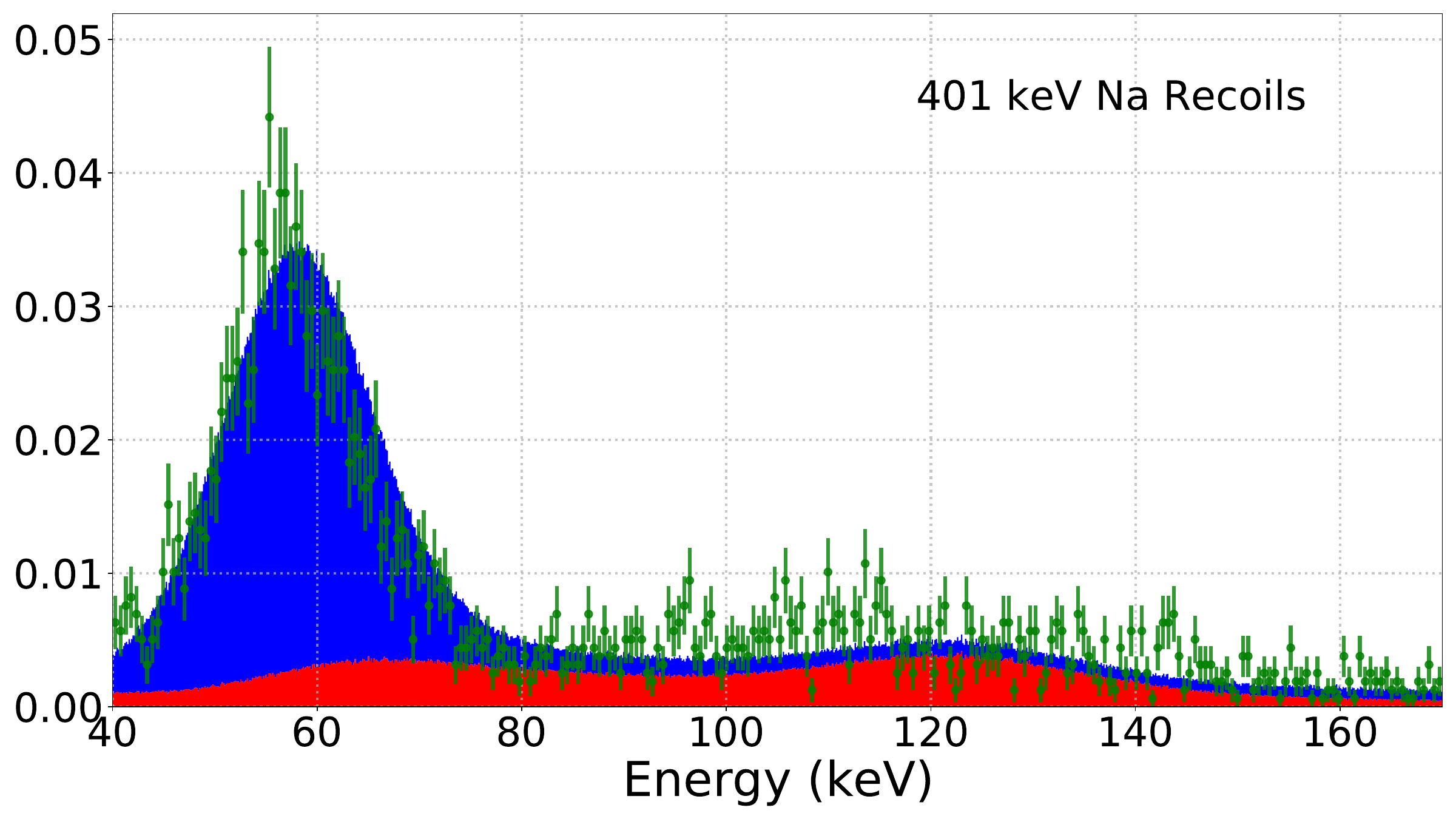}}
        \label{fig:fit5_5}
    \end{subfigure}
    \caption{A selection of spectrum fits, where the fit parameters are taken to be the mean of the posterior distribution given by the Markov Chain Monte Carlo sampling. The red and blue stacked histograms denote the Na and I recoil contributions, respectively, while the green points are the measured spectra. All spectra are normalised to unit area. The Na elastic scattering peak can be seen to move to higher energy and become broader as the recoil energy is increased. Note the different energy scale in the highest recoil energy plots (g) and (h).}
    \label{fig:QFfits_movingElastics}
\end{figure}

Figure \ref{fig:QFfits_movingElastics} shows some of the fitted recoil spectra across the range of Na recoil energies studied. The fits describe the data reasonably well around the recoil peak and the inelastic excitation lines. The low-energy rise is due to the tail of the I elastic scattering peak.

We chose the fit range to be as large as possible to make full use of the measured data, including the inelastic excitation peaks, which carry information about the iodine recoils that would otherwise fall below threshold. An exception is the highest energy fitted Na recoils, associated with the 112.5$^\circ$ detector at 5.2 MeV beam energy, and the 90$^\circ$ detector at 6 MeV beam energy. In these cases the elastic scattering distributions were quite broad and these fits tended to find quenching factors which optimised the Na elastic scattering to other features of the spectrum. In these cases the fit was performed using a restricted energy range to avoid this problem (figures \ref{fig:QFfits_movingElastics}(g) and \ref{fig:QFfits_movingElastics}(h)). 

\begin{figure}
    \centering
    \includegraphics[width=\textwidth]{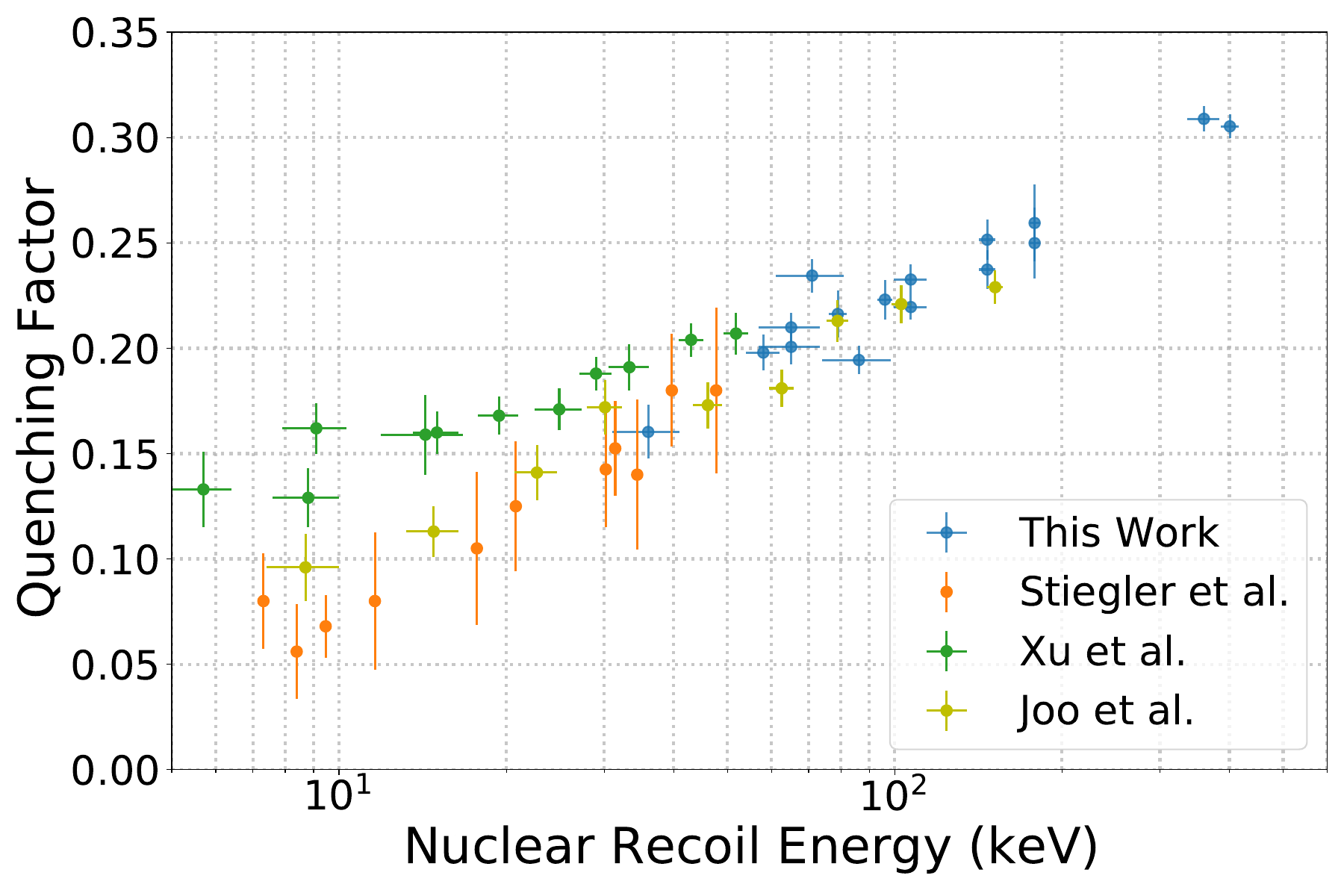}
    \caption{The sodium recoil quenching factors in NaI:Tl inferred from our measurements, together with previous results \cite{Xu2015, stiegler2017, Joo2019}. The quenching factor uncertainties from \cite{stiegler2017} were not given in the paper, and have been inferred from their plot}
    \label{fig:QFplot}
\end{figure}

The Na recoil quenching factors are given in figure \ref{fig:QFplot} and table \ref{tab:QFs}. The Na recoil energies are taken to be the mean of the elastic scattering peak predicted by SimLiT.

Our fits also included the I quenching factor. These were constrained by the location of the main iodine inelastic excitation peak. However, these quenching factors tended to have large uncertainties and are not reported. The covariance between the I and Na quenching factors was not large for any of the fits, so that the poor constraints on the I quenching factors do not strongly influence the Na results.

The systematic errors that could affect our results have been minimised as much as possible. The event selection for tagging nuclear recoil events was deliberately chosen to be relatively conservative to avoid contamination from electron recoil events, and we assume that the small remaining proportion of such events does not affect the quenching factor in a significant way. An imperfect knowledge of the resolution and energy calibration at low energies can also give rise to systematic errors. The resolution model was extrapolated below the energy threshold, which reproduced the observed rise at low energy associated with the tail of the iodine elastic scattering peak. The effect of these sources of systematic error is difficult to assess quantitatively, however we assume such errors are small owing to our relatively high analysis threshold.

\begin{table}[tbh]
    \centering
    \begin{tabular}{m{2.9cm}|m{2.4cm}|m{3.2cm}|m{3.0cm}}
        \hline
        \hline
        Scattered neutron angle & Proton Beam Energy (MeV) & Recoil Energy (keV) & Quenching Factor \\
        \hline
        \hline
        67.5$^\circ$& 2.44  & 36 $\pm$ 5    & 0.160 $\pm$ 0.013\\
        90$^\circ$  & 2.44  & 58 $\pm$ 1    & 0.198 $\pm$ 0.009\\
        67.5$^\circ$& 3     & 65 $\pm$ 8    & 0.201 $\pm$ 0.008\\
        67.5$^\circ$& 3$^{\dagger}$     & 65 $\pm$ 8    & 0.210 $\pm$ 0.007\\
        40$^\circ$  & 5.2   & 71 $\pm$ 10   & 0.235 $\pm$ 0.008\\
        112.5$^\circ$&2.44  & 79 $\pm$ 3    & 0.216 $\pm$ 0.011\\
        40$^\circ$  & 6     & 86 $\pm$ 12   & 0.194 $\pm$ 0.007\\
        135$^\circ$ & 2.44  & 96 $\pm$ 3    & 0.223 $\pm$ 0.009\\
        90$^\circ$  & 3     & 107 $\pm$ 7   & 0.233 $\pm$ 0.007\\
        90$^\circ$  & 3$^{\dagger}$     & 107 $\pm$ 7   & 0.220 $\pm$ 0.006\\
        112.5$^\circ$&3     & 147 $\pm$ 5   & 0.252 $\pm$ 0.010\\
        112.5$^\circ$&3$^{\dagger}$     & 147 $\pm$ 5   & 0.237 $\pm$ 0.009\\
        135$^\circ$ & 3     & 179 $\pm$ 4   & 0.260 $\pm$ 0.018\\
        135$^\circ$ & 3$^{\dagger}$     & 179 $\pm$ 4   & 0.250 $\pm$ 0.017\\
        90$^\circ$  & 6     & 360 $\pm$ 24  & 0.309 $\pm$ 0.006\\
        112.5$^\circ$&5.2   & 401 $\pm$ 15  & 0.305 $\pm$ 0.006\\
    \end{tabular}
    \caption{The measured Na quenching factors at the elastic scattering recoil energies predicted by SimLiT. The measurements with the 3 MeV proton beam were repeated across two different measurement campaigns. The uncertainties are statistical. For the repeated measurements, $^{\dagger}$ indicates results from the second measurement campaign.}
    \label{tab:QFs}
\end{table}

\section{Discussion and Future Work}

Our results are self-consistent across two measurement campaigns and different neutron beam energies. They are also consistent with previous measurements insofar as they show a decreasing quenching factor with recoil energy. There is some tension amongst the previous studies, which may arise from a crystal-dependence to the quenching factor or systematic differences between the methodologies employed in those studies. Our measurement results show a mild preference for the lower quenching factors amongst those previously measured, though we were limited in our ability to examine these differences by the relatively high energy threshold.

The main outcome of this work is to demonstrate a spectrum fitting methodology for the measurement of the Na quenching factors in NaI:Tl with $^{7}$Li(p,n)$^{7}$Be neutrons. Our fits exhibit excellent agreement with the experimental results over a wide energy range. We intend to perform similar measurements using this methodology on high quality NaI:Tl off-cuts from the SABRE experiment's crystals \cite{Antonello2019} in an optimised enclosure and measurement geometry. These measurements will allow a lower energy quenching factor determination and help to probe any crystal dependence.

\section{Acknowledgements}
We would like to thank the technical staff at the ANU Heavy Ion Accelerator Facility, who fabricated detector housings, built the beamline hardware, and carefully aligned the detectors in the correct geometry. We would also like to thank A. Duffy from Swinburne University for the use of his NaI:Tl crystal. This work was supported by the Australian Research Council Discovery Program through project number DP170101675.

\bibliographystyle{JHEP}
\bibliography{library}

\end{document}